\documentclass[11pt]{article}

\usepackage{aas_macros,amsmath,cite,graphicx,mathtools,setspace,subfigure}
\usepackage{comment}
\usepackage[margin=1in,letterpaper]{geometry}
\usepackage[colorlinks=true]{hyperref}
\usepackage[affil-it]{authblk}
\usepackage[svgnames]{xcolor}

\numberwithin{equation}{section}
\newtagform{brackets}{[}{]}

\newcommand{\pd}{\,\partial}

\begin{document}

\title{Consistent  Blandford-Znajek Expansion}
\date{}
\author[1]{Jay Armas\thanks{j.armas@uva.nl}}
\author[2]{Yangyang Cai\thanks{cyysouldancer@email.arizona.edu}}
\author[3]{Geoffrey Comp\`{e}re\thanks{gcompere@ulb.ac.be}}
\author[4]{David Garfinkle\thanks{garfinkl@oakland.edu}}
\author[2]{Samuel E. Gralla\thanks{sgralla@email.arizona.edu}}
\affil[1]{\small University of Amsterdam and Dutch Institute for Emergent Phenomena, 1090 GL Amsterdam, The Netherlands}
\affil[2]{\small Department of Physics, University of Arizona, 1118 E 4th Street, U.S.A}
\affil[3]{\small
Universit\'{e} Libre de Bruxelles and International Solvay Institutes\\
Brussels, CP 231 B-1050, Belgium}
\affil[4]{\small Department of Physics, Oakland University, 146 Library Drive, Rochester, MI 48309, U.S.A }

\maketitle

\begin{abstract}
The Blandford-Znajek mechanism is the continuous extraction of energy from a rotating black hole via plasma currents flowing on magnetic field lines threading the horizon.  In the discovery paper, Blandford and Znajek demonstrated the mechanism by solving the equations of force-free electrodynamics in a perturbative expansion valid at small black hole spin.  Attempts to extend this perturbation analysis to higher order have encountered inconsistencies.  We overcome this problem using the method of matched asymptotic expansions, taking care to resolve all of the singular surfaces (light surfaces) in the problem.  Working with the monopole field configuration, we show explicitly how the inconsistencies are resolved in this framework and calculate the field configuration to one order higher than previously known.  However, there is no correction to the energy extraction rate at this order.  These results confirm the basic consistency of the split monopole at small spin and lay a foundation for further perturbative studies of the Blandford-Znajek mechanism.
\end{abstract}

\vfill\pagebreak

\tableofcontents

\vspace{-.2cm}
\section{Introduction}\vspace{-.2cm}

In 1977 Blandford and Znajek (BZ) \cite{blandford-znajek1977}  discovered that magnetized plasma can continuously extract rotational energy from black holes, even in a steady state configuration.  They proposed this process as the basic engine for relativistic jets from active galaxies, with the magnetic field provided by an accretion disk and the plasma generated self-consistently by particle acceleration and collision in rotation-induced electric fields.  
Large-scale numerical simulations support the basic features of this model \cite{EHTcodeComparison,parfrey-philippov-cerutti2019}, which remains the leading candidate for the power source of relativistic jets. 

The simplest setting for studying the BZ mechanism is the ``split monopole'' configuration, where the accretion disk is replaced by a thin current sheet in the equatorial plane of the black hole, and one considers surrounding force-free plasma with asymptotically radial magnetic field lines.  BZ found an approximate analytic solution valid at small spin \cite{blandford-znajek1977}, and the finite-spin version has been studied numerically \cite{komissarov2001,tchekhovskoy-narayan-mckinney2010,nathanail-contopoulos2014}.  However, attempts to extend the BZ perturbation analysis to higher order have been only partially successful.  Tanabe and Nagataki  \cite{tanabe-nagataki2008} computed the subleading correction to the energy extraction rate, but noted the full subleading solution could not be determined from their method.  More recently, there have been conflicting claims of a high-order solution  \cite{pan-yu2015,pan-yu2015b} and that the expansion is inconsistent  \cite{orselli-etal2018,orselli-etal2019}.

In this paper we introduce an approximation method for the split monopole that resolves all previous difficulties and appears to extend to arbitrary perturbative order and to any magnetic field configuration.  Building on analogous work on the pulsar magnetosphere  \cite{gralla-lupsasca-philippov2016,gralla-lupsasca-philippov2017}, we use the method of matched asymptotic expansions to resolve the important physical scales in the problem.  In particular, a black hole magnetosphere has both an inner and an outer light surface \cite{komissarov2004}, which approach the boundaries of the physical domain (the event horizon and infinity) in the original BZ expansion.  This necessitates the introduction of two additional small-spin expansions (one resolving each light surface), and matching the three expansions together provides the self-consistent solution to the problem, order by order in spin.  Although only one extra expansion (resolving the outer light surface) is needed at the perturbative order we consider, we discuss all three in order to lay a foundation for higher-order studies or studies involving other magnetic field configurations.

We carry out the approach to three perturbative orders higher than leading and explore some aspects of the next correction.  At leading order we reproduce the BZ solution.  At next order all corrections vanish.  At second relative order we confirm the subleading energy extraction rate of Ref.~\cite{tanabe-nagataki2008}.  At third relative order we resolve the issues with previous approaches, showing that the perturbation analysis produces consistent equations with solutions we find numerically.   Interestingly, while the field configuration is corrected at this order, a simple analytic argument shows that the total energy extraction rate is not.  We explore some aspects of the next relative order, finding that logarithms will appear.

Although current astrophysical questions do not demand the calculation of high-order corrections, we have undertaken this calculation on the principle that a fundamental mechanism in relativistic astrophysics deserves a correspondingly thorough treatment.  In addition, the perturbative results may help benchmark high-accuracy codes at small spin where the separation of scales makes numerical work challenging, and the perturbative method may prove useful for other magnetic field calculations.  Finally, by clarifying the consistency of the small-spin expansion, our results should remove any lingering doubts about the consistency of the BZ mechanism itself.

In Sec.~\ref{sec:problem} we formulate the problem at finite spin, and in Sec.~\ref{sec:method} we introduce the perturbation method.  We apply the method in Sec.~\ref{sec:doit} and compare with previous work in  Sec.~\ref{sec:previous}.  We follow the conventions of \cite{gralla-jacobson2014}. 

\section{Statement of the Problem}\label{sec:problem}
We work in Boyer-Lindquist coordinates for the Kerr spacetime, expressing the metric as 
\begin{align}
    ds^2 = - \alpha^2 dt^2 + \rho^2 (d\phi -\Omega_Z dt)^2 + \Sigma \left(\frac{dr^2}{\Delta} + d\theta^2\right),
\end{align}
where
\begin{align}
    \alpha^2 = \Sigma \Delta / A, \qquad 
    \rho^2 = A/\Sigma, \qquad 
    \Omega_Z & = 2 M a r / A 
\end{align}
with \begin{align}
    \Delta & = r^2 - 2Mr + a^2 \\
    \Sigma & = r^2 + a^2 \cos^2  \theta\\
    A & = (r^2+a^2)^2-a^2 \Delta \sin^2\theta.
\end{align}
Note that $\alpha$ and $\Omega_Z$ are the redshift factor and angular velocity of observers with zero angular momentum.  The outer root of $\Delta$ is the event horizon,
\begin{align}
    r_H= M + \sqrt{M^2 - a^2},
\end{align}
which rotates with angular velocity 
\begin{align}
    \Omega_H= \frac{a}{r_H^2+a^2}.
\end{align}
\subsection{Stream Equation}

We now summarize the standard approach to stationary, axisymmetric, force-free fields, using the approach and conventions of \cite{gralla-jacobson2014}. A stationary, axisymmetric, closed two-form $F$ defines a flux function $\psi(r,\theta)$ by
\begin{align}
    \psi(r,\theta) = \frac{1}{2\pi} \int F,
\end{align}
where the integral is over a surface that is bounded by a loop of revolution at $(r,\theta)$, remains outside the black hole, and pierces the northern symmetry axis $\theta=0$ exactly once.  This implies that
\begin{align}\label{psiaxis}
    \psi(r,0)=0, \qquad \psi(r,\pi)=2 \psi_0, 
\end{align}
where $\psi_0$ is the monopole charge of the configuration.  

When the field is furthermore degenerate ($F\wedge F=0$), it defines magnetic field line worldsheets that rotate with an angular velocity $\Omega(\psi)$.  It is useful to introduce a ``co-rotation one-form''
\begin{align}
   \eta=d\phi - \Omega dt, \qquad \qquad |\eta|^2 = \frac{1}{\rho^2} - \frac{(\Omega-\Omega_Z)^2}{\alpha^2}.
\end{align}
Where $|\eta|^2=0$ an observer co-rotating with the field lines would move at the speed of light.  These ``light surfaces'' will play an important role in our analysis.
 
 A field that is furthermore force-free defines a ``polar current'' $I(\psi)$ giving the current flowing through the loop of revolution.  The force-free condition implies that $\psi$ satisfies the ``stream equation'',
\begin{align}\label{stream}
    \nabla_a (|\eta|^2 \nabla^a \psi)+\frac{\Omega'(\Omega-\Omega_Z)}{\alpha^2} \nabla_a \psi \nabla^a\psi +\frac{I I'}{4\pi^2 \alpha^2 \rho^2}=0,
\end{align}
where a prime denotes derivative with respect to $\psi$.  
The requirements that $I=I(\psi)$ and $\Omega=\Omega(\psi)$ may be expressed as
\begin{align}\label{cq}
    dI \wedge d\psi =0, \qquad  d\Omega \wedge d\psi=0.
\end{align}
After solving Eqs.~\eqref{stream} and \eqref{cq} for $\psi$, $I$, and $\Omega$, the field strength may be reconstructed as
\begin{align}\label{F}
    F = \frac{I}{2\pi} \frac{r^2+a^2 \cos^2 \theta}{(r^2+a^2-2Mr) \sin \theta} dr \wedge d\theta + d\psi \wedge \eta.
\end{align}
Note that every stationary, axisymmetric, degenerate, closed two-form with non-zero poloidal magnetic field $(F \cdot \pd_\phi \neq 0)$ can be expressed in this form.  Thus, apart from trivial cases of purely toroidal magnetic field, every stationary, axisymmetric force-free solution can be found by this method.

Surfaces $\psi=\rm{const}$ are called ``poloidal field lines''.  Energy flows only along these surfaces, with the power per field line given by 
\begin{align}\label{dP}
    dP=- I \Omega d\psi.
\end{align}

\subsection{Boundary Conditions}\label{sec:bc}

Eqs.~\eqref{stream} and \eqref{cq} are of a rather non-standard form, and little is known about the boundary value problem in general.  However, actionable understanding has been obtained for the main cases of interest via heuristic arguments and numerical experimentation---see e.g. \cite{gralla-jacobson2014,nathanail-contopoulos2014} for discussion.  For the monopole solution we seek, it appears that the relevant boundary conditions are simply
\begin{enumerate}
    \item The field $F$ is finite on the Kerr exterior and future event horizon.
    \item As $r \to \infty$, the flux $\psi(r,\theta)$ remains bounded and the energy is outgoing ($dP \geq 0$).
\end{enumerate}
The first condition implies that the solution can occur in a black hole formed from collapse (i.e. an astrophysical black hole), while the second corresponds to an isolated black hole (no external source of magnetic field or energy).  

We conjecture that these conditions give rise to a unique solution up to an overall constant factor, which can be taken to be the monopole charge.  This conjecture is consistent with numerical studies and supported by our perturbative analysis.  A proof would in effect generalize the no-hair theorem to include black holes immersed in force-free plasma.\footnote{An astrophysical black hole cannot carry monopole charge, so the theorem would imply that there is no isolated black hole magnetosphere.  This would give rigorous mathematical expression to the well-established idea that an external magnetic field is required to support a black hole magnetosphere.  A uniqueness theorem would also inform discussion of the status of the split monopole as a kind of metastable ground state, decaying according to the lifetime of the current sheet that supports it \cite{lyutikov-mckinney2011}.} 

For use in practice, we now note three consequences of these assumptions.  First, assumption 1 implies a relationship among $I$, $\Omega$, and $\psi$ at the horizon $r=r_H$ (e.g. \cite{gralla-jacobson2014}),
\begin{align}\label{Zhor}
    I=2\pi (\Omega -\Omega_H) \frac{r^2_H+a^2}{r^2_H+a^2\cos^2 \theta}\sin\theta \pd_\theta \psi,\qquad  r=r_H.
\end{align}
This relationship is called the Znajek condition \cite{znajek1977}.  It may be derived by changing to regular coordinates in Eq.~\eqref{F}.  Similarly, assumption 2 implies
\begin{align}\label{Zinf}
    I=-2\pi \Omega \sin \theta \partial_\theta \psi,\qquad  r \to \infty,
\end{align}
 which can be thought of as a Znajek condition at infinity.  It may be derived from assumption 2 by solving the stream equation \eqref{stream} at large $r$ \cite{nathanail-contopoulos2014}.\footnote{We assume a smooth expansion at infinity, $\psi(\theta) = \psi_\infty(\theta)+O(1/r)$, which is consistent with known properties of the solution.}  Alternatively, it follows from regularity conditions at future null infinity \cite{penna2015}.

Finally, regularity of $F$ on the symmetry axis (part of assumption 1) requires that $\pd_\theta \psi$ vanish there.  Combined with Eqs.~\eqref{psiaxis}, we have
\begin{align}\label{psiax}
    \psi|_{\theta=0} = 0, \quad  \psi|_{\theta=\pi} = 2\psi_0, \qquad \pd_\theta \psi|_{\theta=0} = \pd_\theta \psi|_{\theta=\pi}  = 0.
\end{align}

The constant $\psi_0$ fixes the overall normalization and is proportional to the magnetic monopole charge of the configuration.  However, in the astrophysical application we would ``split'' the monopole by multiplying the electromagnetic field $F$ \eqref{F} by an overall factor of $\textrm{sign}(\cos \theta)$. This procedure makes sense because the non-split monopole solution for $F$ is odd under equatorial reflections,\footnote{The equations and boundary conditions for $(\psi,I,\Omega)$ are invariant under $\theta \to \pi - \theta$ and $\psi \to 2 \psi_0 - \psi$, meaning that $d\psi$ flips sign under equatorial reflection.  It then follows from Eq.~\eqref{F} that $F$ is odd under reflection.} so that the split monopole is even, and the magnetic charge is eliminated.  The resulting equatorial discontinuity corresponds to a thin sheet of charge and current that may be viewed as the source of the magnetic field.  In the split monopole $\psi_0$ stands for the total magnetic flux per hemisphere.  In this paper we will always discuss the non-split monopole; the split case follows straightforwardly as described.  In particular, the power radiated is the same for the non-split and the split monopoles.

 Notice that the equations and boundary conditions for $(\psi,I,\Omega)$ are invariant under the simultaneous operation $a \to -a$, $I \to -I$, and $\Omega \to -\Omega$.  This implies in particular that the power \eqref{dP} can depend only on $|a|$.  We will assume $a>0$ without loss of generality. 

\section{Perturbation Method}\label{sec:method}
We now seek a perturbative solution in the dimensionless spin of the black hole,
\begin{align}
    \epsilon = \frac{a}{M}>0.
\end{align}
We can anticipate some non-uniformity in the expansion based on the intuition that important physics occurs near light surfaces, which are horizons for particles moving on field lines  \cite{komissarov2004,gralla-jacobson2014}.  A black hole magnetosphere generically has two such surfaces, which we expect to scale as\footnote{The boundary condition \eqref{Zhor} suggests that $\Omega \sim \Omega_H$.  Choosing $\Omega$ to be a constant proportional to $\Omega_H$ for simplicity, solving $|\eta|^2=0$ yields the scalings shown.}
\begin{align}
    r-r_H & \sim M \epsilon^2, \qquad \textrm{Inner light surface (ILS)} \\
    r & \sim M/\epsilon, \qquad \textrm{Outer light surface (OLS)}. 
\end{align}
As we take $\epsilon \to 0$ fixing $r$ and $M$ (the usual expansion considered previously), the ILS approaches the horizon, while the OLS approaches infinity.  That is, the light surfaces approach the boundaries of the problem.  This perturbative expansion will invariably miss physics occuring on the light surface scales, and its individual terms will (in general) fail to display the boundary behavior of the exact solution.  Additional $\epsilon \to 0$ expansions resolving the ILS and OLS will in general be required to resolve the physics and recover the proper boundary behavior.

In the perturbative solution of the equations, the need for additional expansions is seen directly by the inability to satisfy all boundary conditions within a single expansion.  This difficulty was discovered by Tanabe and Nagatake in 2008 \cite{tanabe-nagataki2008}, who found an inconsistency at fourth order in $\epsilon$.  In 2018 the authors of Ref.~\cite{orselli-etal2018}  showed that there is a problem already at second order, since the approximate solution is not consistent with the exact stream equation expanded at large $r$.  To rectify these problems we consider three distinct expansions: near, mid, and far.\footnote{The problems occur at large $r$, suggesting that only the far expansion is require to cure them.  This is in fact the case, but we will  consider all three limits in order to establish a formalism that should work at any perturbative order, for any magnetic field configuration.}  The near expansion resolves the ILS; the far expansion resolves the OLS; and the mid expansion is the usual one previously considered (which resolves neither). We define characteristic scales associated with each limit,
\begin{align}
    R_{\rm near} = \frac{a^2}{M}, \qquad R_{\rm mid} = M, \qquad  R_{\rm far}  = \frac{M^2}{a},
\end{align}
and introduce associated dimensionless coordinates,
 \begin{align}
     y & = \frac{r-r_H}{R_{\rm near}} = \frac{M(r-r_H)}{a^2} \label{y} \\
     x & = \frac{r}{R_{\rm mid}} = \frac{r}{M} \label{x} \\
     \bar{x} & = \frac{r}{R_{\rm far}} = \frac{a r}{M^2}. \label{xbar}
 \end{align}
We define the near/mid/far expansions as $\epsilon \to 0$ fixing $R_{\rm near}/R_{\rm mid}/R_{\rm far}$ and $y/x/\bar{x}$, where $\theta$ is also fixed in all limits.    Noting that $R_{\rm near}=M \epsilon^2$, $R_{\rm mid}=M$, and $R_{\rm far}=M/\epsilon$, we have a hierarchy of scales $R_{\rm near} \ll R_{\rm mid} \ll R_{\rm far}$ as $\epsilon \to 0$.  We may therefore associate  overlapping regimes of validity to the expansions:
\begin{align}
    \text{near expansion:} \quad & \epsilon \to 0 \  \text{fixed} \ R_{\rm near},y \ \qquad \qquad r-r_H \ll R_{\rm mid}\\
    \text{mid expansion:}\quad  & \epsilon \to 0 \  \text{fixed} \ R_{\rm mid},x \ \quad \ R_{\rm near}\ll r \ll R_{\rm far}\\
    \text{far expansion:} \quad & \epsilon \to 0 \  \text{fixed} \ R_{\rm far},\bar{x} \ \ \ \! \quad \qquad \qquad r \gg R_{\rm mid}
\end{align} 
We introduce order symbols $O_{\rm near}, O_{\rm mid}, \text{and}\ O_{\rm far}$ representing scalings with $\epsilon$ in each limit. Since these can be counter-intuitive, we list some common scalings here:
\begin{subequations}\label{scalings}
 \begin{align}
     M &=O_{\rm near}({\epsilon}^{-2})=O_{\rm mid}(1)=O_{\rm far}(\epsilon)\\
     a &=O_{\rm near}({\epsilon}^{-1})=O_{\rm mid}(\epsilon)=O_{\rm far}({\epsilon}^2)\\
     r_H & = O_{\rm near}({\epsilon}^{-2})=O_{\rm mid}(1)=O_{\rm far}(\epsilon) \\
     \Omega_H &=O_{\rm near}({\epsilon}^{3})=O_{\rm mid}(\epsilon)=O_{\rm far}(1).
 \end{align}
 \end{subequations}
  For expansions in the various limits, we will use a superscript for the order and a subscript for the limit,
 \begin{align}\label{Q}
     Q = Q^{(0)}_{\rm lim} + \epsilon Q^{(1)}_{\rm lim} + \epsilon^2 Q^{(2)}_{\rm lim} + O_{\rm lim}(\epsilon^3).
 \end{align}
 Here $Q$ is any quantity and lim stands for near, mid, or far.  This simple form of the expansion presumes the lack of fractional powers or logs (or worse).  We will see that it suffices at the perturbative order we consider.

 The method of matched asymptotic expansions requires that the various expansions agree in the regimes of overlapping validity.  The near expansion at large $y$ must agree, order by order, with the mid expansion at small $x$.  Similarly, the mid expansion at large $x$ must agree, order by order, with the far expansion at small $\bar{x}$.  One proceeds by guessing the form of the three expansions and modifying as necessary to attain proper matching.  It is difficult to prove uniqueness in such a guess-and-check context, but we can provide some evidence by systematically searching the typical types of expansions.  We will use a ``reluctant log'' strategy, where all expansions are assumed smooth in $\epsilon$ until the presence of logs becomes unavoidable to ensure consistent matching.  We will find that logs are not necessary to resolve previous inconsistencies, but they do appear at one order higher than we consider (App.~\ref{sec:L4}).
 
 \section{Perturbative Solution}\label{sec:doit}
 
 The method outlined in Sec.~\ref{sec:method} above should work for any physical configuration.  In this paper we apply it systematically to the monopole configuration, as defined by the boundary conditions of Sec.~\ref{sec:bc}.  We treat each order in perturbation theory in a separate subsection, in each of which we first state the results and then proceed to derive them.
 
 \subsection{Leading scalings}
 We can establish certain scalings by a general argument.  We first note that the ``no ingrown hair'' theorem \cite{macdonald-thorne1982,gralla-jacobson2014} guarantees that all poloidal field lines ($\psi= {\rm const}$ surfaces) link the horizon to infinity.  Thus the flux function must appear at lowest order in all three limits,
 \begin{align}\label{psiscale}
     \psi = O_{\rm near}(1) = O_{\rm mid}(1) = O_{\rm far}(1).
 \end{align}
 However, let $\theta_H$ denote the angle of intersection of  a poloidal field line with the horizon, and let $\theta_\infty$ denote its angle of intersection with infinity.  Then the Znajek conditions \eqref{Zhor} and \eqref{Zinf} can be solved to yield \cite{penna2015impedance}
 \begin{align}
     \frac{\Omega_H}{\Omega} = -1 - \frac{(\sin \theta \pd_\theta \psi)|_{r=\infty,\theta=\theta_\infty}}{\left.\frac{(r^2_H+a^2)\sin \theta \pd_\theta \psi}{r^2_H+a^2\cos^2 \theta}\right|_{r=r_H,\theta=\theta_H}}.
 \end{align}
 Since the right-hand side is $O(1)$ (in all three limits), we learn that $\Omega$ scales the same way as $\Omega_H$.  From Eq.~\eqref{scalings} we thus have
 \begin{align}\label{Omegascale}
     \Omega = O_{\rm near}(\epsilon^3) = O_{\rm mid}(\epsilon) = O_{\rm far}(1).
 \end{align}
 The Znajek conditions \eqref{Zhor} and \eqref{Zinf} now imply the same scalings for $I$,
 \begin{align}\label{Iscale}
     I = O_{\rm near}(\epsilon^3) = O_{\rm mid}(\epsilon) = O_{\rm far}(1).
 \end{align}
 When possible, we will discuss ``relative order'' instead of these detailed scalings.  For example, the leading order flux function in the far limit refers to $\psi^{(0)}_{\rm far}$, while the leading order polar current in the near limit refers to $\epsilon^3 I^{(3)}_{\rm near}$.  For corrections we discuss the $n{}^{\rm th}$ relative order, defined as $n$ powers of $\epsilon$ higher than the leading behavior.  For example, second relative order refers to $\psi^{(2)}_{\rm far}$, $I^{(5)}_{\rm near}$, and $\Omega^{(3)}_{\rm mid}$.

 \subsection{Leading Order}
 We now determine the leading order behavior of $\psi$, $I$, and $\Omega$ in each of the limits.  The results take an identical form in all limits, but to illustrate the method we present full details: 
 \begin{subequations}\label{leading}
\begin{align}
\psi^{(0)}_{\rm near} & = \psi^{(0)}_{\rm mid} = \psi^{(0)}_{\rm far} = \psi_0(1-\cos \theta) \\
\epsilon^3 \Omega^{(3)}_{\rm near} & = \epsilon \Omega^{(1)}_{\rm mid} = \Omega^{(0)}_{\rm far} = \frac{a}{8M^2}. \\
    \epsilon^3 I^{(3)}_{\rm near} & = \epsilon I^{(1)}_{\rm mid} = I^{(0)}_{\rm far} = - 2 \pi \frac{a}{8 M^2} \psi_0 \sin^2 \theta.
\end{align}
\end{subequations}
This reproduces the original Blandford-Znajek result in our formal scheme.  The total power may be computed from \eqref{dP} using any of the three limits.  We will use the far limit so that the energy extraction occurs at zeroth order.  One finds\footnote{If one prefers to consider the split monopole, the integral is replaced by twice the integral from zero to $\pi/2$, giving the same result.}
\begin{align}\label{P0}
 P^{(0)}_{\rm far} = 2\pi \left( \frac{a}{8M} \right)^2 \int_0^\pi \sin^3 \! \theta \ \! d\theta = \frac{\pi}{24} \frac{a^2 \psi_0^2}{M^4}.
\end{align}
This is the famous electromagnetic extraction of energy \cite{blandford-znajek1977}.  Noting that $\Omega^{(0)}_{H,{\rm far}}=a/(4M^2)$, we may also write
\begin{align}\label{leading-simple}
    \psi \approx \psi_0(1-\cos \theta), \qquad \Omega \approx \frac{\Omega_H}{2}, \qquad I \approx -2\pi \frac{\Omega_H}{2} \psi_0 \sin ^2 \theta, \qquad     P \approx \frac{2\pi}{3} \psi_0^2 \Omega_H^2,
\end{align}
where $\approx$ means valid to leading order in any of the limits. Using $M \Omega_H$ as the perturbation parameter performs better in comparisons to numerical results at finite spin \cite{TMN2008}, but we will continue to use $\epsilon=a/M$ for simplicity; the results are easily converted if desired.  We now derive Eqs.~\eqref{leading} by working to leading order in all three limits.

\subsubsection{Mid limit}
We begin with the mid limit.  Since $I$ and $\Omega$ are $O_{\rm mid}(\epsilon)$, these quantities do not appear in the stream equation \eqref{stream} at leading order.  Instead, we have the simple linear equation
\begin{align}\label{mideq0}
    L_{\rm mid}[\psi^{(0)}_{\rm mid}]=0,
\end{align}
where in terms of $x=r/M$ we have
 \begin{align}
    L_{\rm mid}={\partial}_x \left(\left[1-\frac{2}{x}\right]{\partial}_x\right)+\frac{\sin\theta}{x^2}{\partial}_\theta\left(\frac{1}{\sin\theta}{\partial}_\theta\right).
\end{align}
This is the stream equation in Schwarzschild spacetime with $I=\Omega=0$, which corresponds to stationary, axisymmetric vacuum magnetic fields in Schwarzschild.  The general solution satisfying the conditions \eqref{psiax} at the poles is given in a multipole expansion as (App.~\ref{sec:eigen}),
\begin{align}\label{psi0midsoln}
    \psi^{(0)}_{\rm mid} = \psi_0(1-\cos \theta) + \sum_{\ell=1}^\infty \left( B^<_\ell R^<_\ell(x) + B^>_\ell R^>_\ell(x) \right) \Theta_\ell(\theta),
\end{align}
where $B_\ell^<$ and $B_\ell^>$ are numbers.  The boundary conditions are a smooth match to the near and far expansions, which requires finiteness of $\psi^{(0)}_{\rm mid}$ as $x\to 2$ and $x \to \infty$.\footnote{Changing to near coordinates $y=\epsilon^{-2} (x-r/r_H) \approx (x-2)/\epsilon^2$, we see that any term in the mid expansion that blows up as $x\to 2$ will require a term in the near expansion that blows up as $\epsilon \to 0$, which violates the scalings \eqref{psiscale}.  Changing to far coordinates $\bar{x}=\epsilon x$, we see that any term in the mid expansion that blows up as $x \to \infty$ would similarly require a term in the far expansion that blows up as $\epsilon \to 0$.}  This sets $B_\ell^<=B_\ell^>=0$, since $R^>_\ell$ blows up as $x \to 2$, while $R^<_\ell$ blows up as $x \to \infty$.  Thus the unique solution is simply
\begin{align}\label{psi0mid}
   \psi^{(0)}_{\rm mid}=\psi_0(1-\cos\theta),
\end{align}
i.e., a pure monopole.

 \subsubsection{Far limit}
  As all quantities are $O(1)$ in the far limit, Eqs.~\eqref{stream} and \eqref{cq} retain their form as $\epsilon \to 0$, with the only modification being that $a$ and $M$ are set to zero.  That is, the far limit of the problem is just the stream equation in flat spacetime.  The boundary conditions are assumption 2 (finite flux at infinity and outgoing energy flux) as well as matching to the mid region \eqref{psi0mid},
\begin{align}\label{mono}
    \psi^{(0)}_{\rm far} \sim \psi_0(1-\cos\theta), \qquad \bar{x} \to 0.
\end{align}
That is, we consider an isolated magnetosphere that becomes monopolar near the origin.  (Note that \eqref{mono} represents a singularity at the origin---a point monopole---since the flux does not vanish there.) This case was analyzed by Michel \cite{michel1973mon}, who found a large class of solutions,
\begin{subequations}\label{Michel}
\begin{align}
    \psi^{(0)}_{\rm far} &=\psi_0(1-\cos\theta), \\
    \Omega^{(0)}_{\rm far}&=\frac{a}{M^2}\omega_0(\theta), \\ I^{(0)}_{\rm far}&=-2\pi\frac{a}{M^2}\omega_0(\theta)\psi_0\sin^2\theta,
\end{align}
\end{subequations}
where $\omega_0(\theta)$ is a free function that was made dimensionless using the far lengthscale $R_{\rm far}=M^2/a$.  

Although we give no rigorous proof, we expect that the solution family \eqref{Michel} is unique, i.e., there are no further solutions satisfying the boundary conditions.  This is supported by numerical experience (e.g.,\cite{contopoulos-kazanas-fendt1999,nathanail-contopoulos2014}) suggesting the following general picture: (1) One may choose the value of $\psi$ on one inner boundary surface (such as a neutron star), but at infinity (and on event horizons) it must be left free; (2) Each light surface in the problem contributes a matching condition that fixes one free function ($I(\psi)$, $\Omega(\psi)$ or some combination).  Here the condition \eqref{mono} is equivalent to choosing $\psi$ on an arbitrarily small sphere, the value at infinity is left free (assumption 2 in Sec.~\ref{sec:bc} above), and the far equation (the stream equation in flat spacetime) has only a single light surface, so we expect to fix only one of the two free functions $I(\psi)$ and $\Omega(\psi)$.  Thus we expect the general solution to have a single free function, as in Eqs.~\eqref{Michel}.

\subsubsection{Near limit}

 Letting $\psi=\psi_{\rm near}^{(0)} + O_{\rm near}(\epsilon)$, $I=\epsilon^3 I_{\rm near}^{(3)} + O_{\rm near}(\epsilon^4)$, and $\Omega=\epsilon^3 \Omega_{\rm near}^{(3)} + O_{\rm near}(\epsilon^4)$ in accordance with the scalings \eqref{psiscale}, \eqref{Omegascale} and \eqref{Iscale}, we find the stream equation \eqref{stream} at leading order in the near limit to be
 \begin{align}\label{horse}
    (\csc^2\theta+2(1-4\hat{\Omega})\partial_y \hat{\Omega})\partial_y\psi^{(0)}_{\rm near}
    +(-\frac{1}{2}+y \csc^2\theta +4\hat{\Omega}-8\hat{\Omega}^2)\partial^2_y\psi^{(0)}_{\rm near}=0,
\end{align}
where we use $y=(r-r_H)/R_{\rm near}$ and introduce
\begin{align}
    \hat{\Omega}(y,\theta) = R_{\rm near} \Omega^{(3)}_{\rm near},
\end{align}
recalling that $R_{\rm near}=a^2/M$.  In obtaining Eq.~\eqref{horse} we have used Eq.~\eqref{cq} to replace $\pd_\theta \hat{\Omega}$ by $\pd_y \hat{\Omega} \pd_\theta \psi^{(0)}_{\rm near}/\pd_y \psi^{(0)}_{\rm near}$.  All $\theta$-derivatives have dropped out, so we in effect have an ordinary differential equation in $y$.  The solution for $\psi^{(0)}_{\rm near}$ must match the mid solution \eqref{psi0mid} at large $y$,
\begin{align}
    \psi^{(0)}_{\rm near} \sim \psi_0(1-\cos \theta), \qquad y \to \infty.
\end{align}
A simple solution is just
\begin{align}\label{psi0near}
    \psi^{(0)}_{\rm near} = \psi_0(1-\cos \theta).
\end{align}
We do not prove uniqueness here.

\subsubsection{Znajek condition}

Since $\psi^{(0)}$ depends only on $\theta$ in all three regions (in fact it is identical), the leading order $I$ and $\Omega$ must also pure functions of $\theta$ in all three regions.  (This follows from $I=I(\psi)$ and $\Omega=\Omega(\psi)$, or more formally from Eqs.~\eqref{cq}.)  Thus the matching is exact,
\begin{align}
\epsilon^3 \Omega^{(3)}_{\rm near} & = \epsilon \Omega^{(1)}_{\rm mid} = \Omega^{(0)}_{\rm far} = \frac{a}{M^2} \omega_0(\theta). \label{omegalead} \\
    \epsilon^3 I^{(3)}_{\rm near} &= \epsilon I^{(1)}_{\rm mid} = I^{(0)}_{\rm far} = - 2 \pi \frac{a}{M^2} \psi_0 \sin^2 \theta \omega_0(\theta).\label{Ilead}
\end{align}
In the near expansion we may apply the Znajek condition \eqref{Zhor} at the horizon.  This equation is non-trivial first at $\epsilon^3$, where it states 
\begin{align}
    I^{(3)}_{\rm near}(\theta)=2\pi \left(\Omega^{(3)}_{\rm near}(\theta) - \frac{M}{4a^2}\right) \sin\theta \pd_\theta \psi^{(0)}_{\rm near}(\theta).
\end{align}
(We need not evaluate at the horizon explicitly, since all quantities above depend only on $\theta$.)  Plugging in using Eqs.~\eqref{psi0near}, \eqref{omegalead} and \eqref{Ilead} then fixes the free function as $\omega_0(\theta)$ to be a constant,
\begin{align}
    \omega_0 = \frac{1}{8}.
\end{align}
This completes the solution of the problem at leading order [Eqs.~\eqref{leading} above].

\subsection{First relative order}

We now proceed to the next relative order in $\epsilon$.  We will find that all quantities vanish, 
\begin{align}\label{vanishparty}
    \psi^{(1)}_{\rm near}  = \psi^{(1)}_{\rm mid} = \psi^{(1)}_{\rm far} = \Omega^{(4)}_{\rm near} = \Omega^{(2)}_{\rm mid} = \Omega^{(1)}_{\rm far} = I^{(4)}_{\rm near} = I^{(2)}_{\rm mid} = I^{(1)}_{\rm far} = 0.
\end{align}
The correction to the power then vanishes as well,
\begin{align}\label{P1}
    P^{(1)}_{\rm far} = 0.
\end{align}
That is, the leading results are not corrected at first relative order in $\epsilon$.  Although we could straightforwardly check that all equations are satisfied, we will instead proceed systematically in order to elucidate the method and inform the question of uniqueness.

In the mid limit, the current and angular velocity are $O(\epsilon)$ and hence do not contribute to the stream equation until $O(\epsilon^2)$.  Thus the mid equation is identical to the leading order equation,
\begin{align}\label{mid1}
    L_{\rm mid}[\psi^{(1)}_{\rm mid}]=0
\end{align}
The general solution for $\psi^{(1)}_{\rm mid}$ is the right-hand-side of \eqref{psi0midsoln} without the $\psi_0$ term, since $\psi^{(1)}_{\rm mid}$ must vanish at both poles to preserve the conditions \eqref{psiax}.  However, the remaining terms are also disallowed by matching to the other expansions,\footnote{Since $R^< \sim x^{\ell+1}$ as $x \to \infty$, it contributes a term of order $\epsilon x^{\ell+1}$ to the expansion of $\psi$.  This reads $\epsilon^{-\ell} \bar{x}^{\ell+1}$ in the far coordinate $\bar{x}=\epsilon x$, and hence would require an inverse power of $\epsilon$ in the far expansion, which we do not allow.  Similarly, $R^> \sim \log(x-2)$ as $x \to 2$, contributing $\epsilon \log(x-2)$ to the expansion.  This scales as $\epsilon \log (\epsilon^2 y)$ in the near coordinate $y\approx(x-2)/\epsilon^2$, generating an $\epsilon \log \epsilon$ term in the near expansion, which is excluded by the reluctant log strategy (see discussion at the conclusion of Sec.~\ref{sec:method}.)} and the only solution is vanishing $\psi^{(1)}_{\rm mid}$,
\begin{align}\label{psi1mid}
\psi^{(1)}_{\rm mid} = 0.
\end{align}
At first relative order in the near limit, we find
\begin{align}
     L_{\rm near}[\psi^{(1)}_{\rm near}]=0
\end{align}
with
\begin{align}\label{Lnear}
      L_{\rm near} = 16\partial_y +(-1+16y+\cos2\theta)\partial^2_y.
\end{align}
The general solution is
  \begin{align}
     \psi^{(1)}_{\rm near}=a(\theta)+b(\theta) \log (-1+16y+\cos 2\theta),
 \end{align}
 for arbitrary functions $a(\theta)$ and $b(\theta)$.  However, the second term blows up on the poles when $y=0$ (the horizon), violating the conditions \eqref{psiax}.  Thus the general solution is a free function $a(\theta)$. 
  
However, this function must vanish to match the vanishing mid solution, so we also have $\psi^{(1)}_{\rm near}=0$.  That is, the only allowed solutions are
\begin{align}
    \psi^{(1)}_{\rm mid} = \psi^{(1)}_{\rm near} = 0. \label{nearmid1}
\end{align}
Eq.~\eqref{cq} now implies that the current and angular velocity only depend on $\theta$ in this regime,
\begin{align}\label{Iomega1mid}
    \epsilon^2 \Omega^{(4)}_{\rm near} = \Omega^{(2)}_{\rm mid} = \frac{1}{M} \omega_1(\theta), \qquad \epsilon^2 I^{(4)}_{\rm near} = I^{(2)}_{\rm mid} = \frac{\psi_0}{M} i_1(\theta),
\end{align}
where $\omega_1$ and $i_1$ are dimensionless.  The Znajek condition at the horizon \eqref{Zhor} may be applied in the near limit.  This yields the restriction
\begin{align}\label{i1}
    i_1(\theta)=2\pi \sin ^2 \theta \omega_1(\theta).
\end{align}

In the far limit $I$, $\Omega$, and $\psi$ all appear at the same order.  Using Eq.~\eqref{cq} and matching to Eqs.~\eqref{psi1mid} and \eqref{Iomega1mid}, we learn that 
\begin{align}\label{I3midcool}
    \Omega^{(1)}_{\rm far} & = \frac{a}{M^2} \omega_1(\theta) \\ I^{(1)}_{\rm far} & =  \frac{a}{M^2} \left( \frac{-\pi}{2} \psi^{(1)}_{\rm far} \cos \theta + \psi_0 i_1(\theta) \right) \\
    \psi^{(1)}_{\rm far} & \sim 0, \quad \bar{x} \to 0.\label{psi1bc}
\end{align}

Plugging in to the stream equation \eqref{stream}, we find 
\begin{align}\label{fareq1}
    L_{\rm far}[\psi_{\rm far}^{(1)}] - \frac{\psi_0}{16\pi \sin \theta} \pd_\theta \left(\sin^2 \theta g_1(\theta) \right)=0,
\end{align}
where
\begin{align}\label{Lfar}
    L_{\rm far} & = \sin \theta \pd_\theta \left( \sin \theta \chi^2 \pd_\theta \right) + \sin^2 \theta  \pd_{\bar{x}} \left( \bar{x}^2 \chi^2 \pd_{\bar{x}} \right) + \frac{1}{32} \left( 2 - 3 \sin^2 \theta\right),\\
    & \qquad \chi^2  =  \frac{1}{\bar{x}^2 \sin ^2 \theta}-\frac{1}{64}.
\end{align}
and
\begin{align}\label{g1}
     g_1(\theta)= i_1(\theta) + 2\pi \sin^2 \theta \omega_1(\theta) = 4\pi \sin^2 \theta \omega_1 (\theta),
\end{align}
where we use \eqref{i1} in the last step.

The boundary conditions are that $\psi^{(1)}$ vanishes at the origin $\bar{x}=0$ [Eq.~\eqref{psi1bc}] and is finite as $\bar{x} \to \infty$ (to continue to satisfy assumption 2).  A simple solution satisfying these conditions is simply $\psi^{(1)}_{\rm far}=0$ with $\omega_1(\theta)=0$.  We expect that this solution is unique by similar reasoning to that given at leading order in the far zone below Eqs.~\eqref{Michel}: we have specified $\psi^{(1)}_{\rm far}$ at small $\bar{x}$, we have left it free at infinity, and the differential equation \eqref{fareq1} contains a single light surface $\bar{x} \sin \theta=8$ that is expected to fix the free function $\omega_1(\theta)$.

Independently of the question of uniqueness, the solution $\psi^{(1)}_{\rm far}=0$ and $\omega_1(\theta)=0$ also requires $i_1(\theta)=0$ by \eqref{i1}, completing the derivation of Eqs.~\eqref{vanishparty}.

\subsection{Second relative order}

At second relative order we find non-trivial corrections.    For the flux function we will find 
\begin{align}\label{subleading-psi}
\psi^{(2)}_{\rm near}  = \psi_0 \hat{R}(2) \sin^2 \theta \cos \theta, \qquad \psi^{(2)}_{\rm mid} & = \psi_0 \hat{R}(x) \sin^2 \theta \cos \theta, \qquad \psi^{(2)}_{\rm far} = 0,
\end{align}
where $\hat{R}(x)$ is a function that vanishes at large $x$, given below in \eqref{Rhat}.  For the rotation we find,
\begin{align}
    \epsilon^3 \Omega^{(5)}_{\rm near} =
    \epsilon\Omega^{(3)}_{\rm mid} = \Omega^{(2)}_{\rm far} = \frac{a}{M^2}\omega_2(\theta),
\end{align}
where $\omega_2(\theta)$ is given below in Eq.~\eqref{omega2}. Finally, the current is given by 
\begin{align}
    \epsilon^3 I^{(5)}_{\rm near} & = -2\pi \psi_0 \frac{a}{M^2} \sin^2 \theta \left( \omega_2(\theta)  + \frac{1}{4} \hat{R}(2)\cos^2 \theta \right), \\
    \epsilon I^{(3)}_{\rm mid} & = -2\pi \psi_0 \frac{a}{M^2} \sin^2 \theta \left( \omega_2(\theta)  + \frac{1}{4} \hat{R}(x)\cos^2 \theta \right), 
    \\ I^{(2)}_{\rm far} & = -2\pi \psi_0 \frac{a}{M^2} \sin^2 \theta \omega_2(\theta).
\end{align}
 Notice that the near and far expansions are still essentially trivial, just reproducing the asymptotic value of the mid function.  This is why previous work using a single expansion did not encounter any inconsistency at this order.  The correction to the energy flux is 
 \begin{align}
     P^{(2)}_{\rm far} = \frac{\pi(56-3\pi^2)}{1080} \frac{a^2 \psi_0^2}{M^4}.
 \end{align}
 These expressions have all appeared before in the literature \cite{blandford-znajek1977,tanabe-nagataki2008,pan-yu2015}; see Sec.~\ref{sec:previous} for detailed discussion.  We now derive these results.

\subsubsection{Mid limit}

We again begin in the mid expansion, where the flux function is decoupled from $I$ and $\Omega$ order by order in the expansion.  At second order, we now find a source term,
\begin{align}\label{mideq2}
    L_{\rm mid}[\psi^{(2)}_{\rm mid}] = s^{(2)}_{\rm mid},
\end{align}
with 
\begin{align}\label{s2mid}
s^{(2)}_{\rm mid}=-\psi_0\frac{(2+x)}{x^4} \Theta_2(\theta),
\end{align}
where we remind the reader that $\Theta_2(\theta)=\cos\theta\sin^2\theta$.  We require that $\psi^{(2)}_{\rm mid}$ vanish at both poles to preserve the conditions \eqref{psiax}.  Decomposing into the  eigenmodes $\Theta_\ell$, the general such solution is
\begin{align}\label{psi2gen}
    \psi^{(2)}_{\rm mid}= \psi_0 \hat{R}(x) \Theta_2(\theta) +  \sum_{\ell=1}^\infty \left( B^<_\ell R^<_\ell(x) + B^>_\ell R^>_\ell(x) \right) \Theta_\ell(\theta),
\end{align}
where $\psi_0 \hat{R}$ is any particular solution to the $\ell=2$ equation.  A particular solution may be found by variation of parameters as 
\begin{align}\label{varyit}
    \psi_0 \hat{R}&=-2R^{<}_2(x) \int R^{>}_2(x)s^{(2)}_{\rm mid}(x) dx + 2R^{>}_2(x) \int R^{<}_2(x)s^{(2)}_{\rm mid}(x)dx,
\end{align}
 Performing the integrals and choosing the free constants for regularity of $\hat{R}$ at $x=2$ and $x=\infty$, we have
\begin{align}\label{Rhat}
    \hat{R}(x)&  = \frac{1}{72x}\Bigg[24+11x+36x^2-36x^3+(6x+18x^2-36x^3)\log\left(\frac{x}{2}\right) \nonumber \\
    & +(27x^3-18x^4)\log\left(\frac{x}{2}\right)\log\left(\frac{x-2}{x}\right)+9x^3(-3+2x)\textrm{Li}_2\left(\frac{2}{x}\right) \Bigg],
\end{align}
where the dilog $\textrm{Li}_2$ is defined as
\begin{align}
    \textrm{Li}_2(z)=- \int^{1}_0 \frac{\log(1-zt)}{t}dt.
\end{align}
The asymptotic behavior is
\begin{subequations}\label{Rhatasy}
\begin{align}
    \hat{R}(2) & = \frac{1}{72}(-49+6\pi^2),  \label{Rhat2}  \\
    \hat{R} & \sim \frac{1}{4x}, \qquad x \to \infty. \label{Rhatinfty}
\end{align}
\end{subequations}

Eq.~\eqref{psi2gen} with Eq.~\eqref{Rhat} constitute the general solution satisfying the pole conditions \eqref{psiax}.  However, matching to the near and far regions again sets $B_\ell^<=B_\ell^>=0$.\footnote{Since $R^< \sim x^{\ell+1}$ as $x \to \infty$, it contributes a term of order $\epsilon^2 x^{\ell+1}$ to the mid expansion at large $x$.  This reads $\epsilon^{-\ell+1} \bar{x}^{-\ell}$ in the far coordinate $\bar{x}=\epsilon x$, which has an illegal inverse power of $\epsilon$ except when $\ell=1$.  In the $\ell=1$ case we have
$\epsilon^{0} \bar{x}$, which would match to a term that grows like $\bar{x}$ at small $\bar{x}$ in the far expansion at zeroth order.  However, such a term is not present in the solution $\psi^{(0)}_{\rm far}=\psi_0(1-\cos \theta)$.  For the other boundary $x \to 2$, we note that $R^> \sim \log(x-2)$ as $x \to 2$, contributing $\epsilon^2 \log(x-2)$ to the mid  expansion.  This reads $\epsilon^2 \log (\epsilon^2 y)$ in the near coordinate $y=(x-2)/\epsilon^2$, generating an illegal $\epsilon^2 \log \epsilon$ term in the near expansion.}  Thus the unique solution is
\begin{align}\label{psi2mid}
    \psi^{(2)}_{\rm mid} = \psi_0 \hat{R}(x) \Theta_2(\theta).
\end{align}
This solution provides boundary conditions for the near and far expansions via matching its asymptotic behavior.  As this is the first order where position-dependence appears, we will devote more explanation to the matching.  At small $x$ we have
\begin{align}
    \psi  \sim \psi_0(1-\cos \theta) + \epsilon^2 \psi_0 \hat{R}(2) \Theta_2(\theta), \qquad \epsilon \stackrel{\rm \tiny mid}{\to} 0 \textrm{ then } x \to 2.
\end{align}
The method of matched asymptotic expansions demands agreement  with the large-$y$ behavior of the near limit.  Thus to achieve a match we require
\begin{align}\label{nearmatch2}
    \psi^{(2)}_{\rm near} \sim \psi_0 \hat{R}(2) \Theta_2(\theta), \qquad y \to \infty.
\end{align}
To understand the role of the large-$x$ behavior, we note that
\begin{align}\label{cute}
    \psi & \sim \psi_0(1-\cos \theta) +\epsilon^2 \psi_0 \frac{1}{4x} \Theta_2(\theta), \qquad  \epsilon \stackrel{\rm \tiny mid}{\to} 0 \textrm{ then } x \to \infty.
\end{align}
This must agree with the small-$\bar{x}$ behavior of the far limit.  Noting $\bar{x} = x\epsilon$, we therefore need
\begin{align}\label{farmatch2}
    \psi^{(2)}_{\rm far} & \sim 0, \qquad \bar{x} \to 0.
\end{align}
This is all that is required for the second relative order, but note that \eqref{cute} also imposes a condition at the next order,
\begin{align}\label{newthing}
    \psi^{(3)}_{\rm far} & \sim \frac{\psi_0}{4\bar{x}} \Theta_2(\theta), \qquad \bar{x} \to 0.
\end{align}
We display this condition here because of its importance in resolving previous inconsistencies.

We finally consider the form of $I$ and $\Omega$.  From Eq.~\eqref{cq} we have
\begin{align}
\Omega^{(3)}_{\rm mid} & = \frac{1}{M} \omega_2(\theta),\label{Omega3mid} \\
    I^{(3)}_{\rm mid} & = \frac{-\pi}{2M} \psi^{(2)}_{\rm mid} \cos \theta + \frac{\psi_0}{M} i_2(\theta),\label{I3mid}
\end{align}
where $\omega_2(\theta)$ and $i_2(\theta)$ are dimensionless.

\subsubsection{Near limit}

At second relative order in the near limit, we again find
\begin{align}
     L_{\rm near}[\psi^{(2)}_{\rm near}]=0,
\end{align}
where $L_{\rm near}$ was given in Eq.~\eqref{Lnear} above.  As before, the general solution is a free function of $\theta$.  This must match the small-$x$ value of the mid limit by  Eq.~\eqref{nearmatch2}, so we have
\begin{align}
    \psi^{(2)}_{\rm near} = \psi_0 \hat{R}(2) \Theta_2(\theta).
\end{align}
Eq.~\eqref{cq} implies that $I$ and $\Omega$ depend only on $\theta$ at this order, and matching to the mid limit \eqref{Omega3mid}-\eqref{I3mid} gives

\begin{align}
    \Omega^{(5)}_{\rm near} & =\epsilon^{-2} \Omega^{(3)}_{\rm mid}= \frac{M}{a^2} \omega_2(\theta) \\
    I^{(5)}_{\rm near} & =\epsilon^{-2} I^{(3)}_{\rm mid}\vert_{x=2} = \frac{-\pi M}{2a^2} \psi_0 \hat{R}(2) \Theta_2(\theta) \cos \theta + \frac{\psi_0 M}{a^2} i_2(\theta).
\end{align}

  The horizon Znajek condition \eqref{Zhor} then yields
\begin{align}\label{i2}
    i_2(\theta) & 
     = 2\pi  \sin^2 \theta \left(\omega_2(\theta)-\frac{1}{16}+\frac{4\hat{R}(2)-1}{32} \sin^2 \theta \right).
\end{align}

\subsubsection{Far limit}
From \eqref{cq} in the far limit at $O(\epsilon^2)$ we find
\begin{align}
    \Omega^{(2)}_{\rm far} & = \frac{a}{M^2} \omega_2(\theta) \\ I^{(2)}_{\rm far} & =  \frac{a}{M^2} \left( \frac{-\pi}{2} \psi^{(2)}_{\rm far} \cos \theta + \psi_0 i_2(\theta) \right),
\end{align}
where we have used \eqref{farmatch2}, \eqref{Omega3mid}, and \eqref{I3mid} to match to the mid limit.  The far equation takes an identical form to the first order equation \eqref{fareq1},
\begin{align}\label{fareq2}
    L_{\rm far}[\psi_{\rm far}^{(2)}] -  \frac{\psi_0}{16\pi \sin \theta} \pd_\theta \left(\sin^2 \theta g_2(\theta) \right) = 0,
\end{align}
with
\begin{align}\label{g2}
     g_2(\theta) &= i_2(\theta) + 2\pi \sin^2 \theta \omega_2(\theta) \\
     & =  4\pi  \sin^2 \theta \left(\omega_2(\theta)-\frac{1}{32}+\frac{4\hat{R}(2)-1}{64} \sin^2 \theta \right),
\end{align}
where we use \eqref{i2} in the second line.  As in first order, Eq.~\eqref{fareq2} has a single light surface and a single free function $\omega_2(\theta)$, so we expect a unique solution for $\psi^{(2)}_{\rm far}$ and $\omega_2$ given the boundary condition \eqref{farmatch2}.  One simple solution is where $\psi^{(2)}_{\rm far}$ and $g_2(\theta)$ both vanish, giving
\begin{align}
    \psi^{(2)}_{\rm far}=0
\end{align}
and 
\begin{align}\label{omega2}
    \omega_2(\theta)=\frac{1}{32}-\frac{4\hat{R}(2)-1}{64} \sin^2 \theta.
\end{align}
As a consistency check, notice that Eqs.~\eqref{i2} and \eqref{omega2} imply
\begin{align}
    i_2(\theta) = -2\pi \omega_2(\theta) \sin^2 \theta,
\end{align}
which is equivalent to the infinity Znajek condition  \eqref{Zinf} given $\psi^{(2)}_{\rm far}=0$.

\subsection{Third relative order}
At third relative order, we find
\begin{align}
    \psi^{(3)}_{\rm near} = \psi^{(3)}_{\rm mid} = 0, \qquad \psi^{(3)}_{\rm far} \neq 0.
\end{align}
We derive the partial differential equation obeyed by $\psi^{(3)}_{\rm far}$ and solve it numerically (Fig.~\ref{psi} below).  The angular velocity and current are given in the far region as
\begin{align}
\Omega^{(3)}_{\rm far}  = \frac{a}{M^2} \omega_3(\theta), \qquad
I^{(3)}_{\rm far}  = - \frac{\pi}{2} \frac{a}{M^2} \psi^{(3)}_{\rm far} \cos \theta + \frac{a}{M^2} 2\pi \psi_0 \sin^2 \theta \omega_3(\theta),
\end{align}
where $\omega_3(\theta)$ is known numerically [Eq.~\eqref{omega3_full} below].  The contribution to the total energy flux turns out to vanish at this order,
\begin{align}
    P^{(3)}_{\rm far} = 0.
\end{align}
Mathematically, this result follows from an integrand being a total derivative (Eq.~\eqref{P3} below); we have not identified any deeper reason for the vanishing of this correction.

\subsubsection{Mid and near limits}

At third relative order in the mid limit, we have no source term (like the first relative order),
\begin{align}
    L_{\rm mid}[\psi^{(3)}_{\rm mid}] = 0.
\end{align}
The general solution for $\psi^{(3)}_{\rm mid}$ is again the right-hand-side of \eqref{psi0midsoln} without the $\psi_0$ term.  However, as before we may exclude the homogeneous terms based on their failure to match to the near and far expansions,\footnote{The argument is identical for $R^>_\ell$, which would introduce $\epsilon^3 \log \epsilon$ terms into the near expansion.  The $R^<_\ell$ terms behave as $x^{\ell+1}$ at large $x$, and hence contribute $\epsilon^3 x^{\ell+1} = \epsilon^{2-\ell} \bar{x}^{\ell+1}$ to the expansion.  The $\ell>2$ terms would introduce invserse powers of $\epsilon$ to the far expansion, while the $\ell=1$ and $\ell=2$ terms do not match the first and zeroth order far expansions (respectively), which have already been determined.} and the only solution is trivial,
\begin{align}\label{psi3mid}
    \psi^{(3)}_{\rm mid}=0.
\end{align}
In the near limit at third relative order we again find
\begin{align}
    L_{\rm near}[\psi^{(3)}_{\rm near}] = 0,
\end{align}
and again the only regular solution is a free function of $\theta$.  To match to the mid solution \eqref{psi3mid} this function must vanish,
\begin{align}
    \psi^{(3)}_{\rm near}=0.
\end{align}

Eq.~\eqref{cq} implies that $I$ and $\Omega$ are just functions of $\theta$ at this order, which match as
\begin{align}\label{Iomega2mid}
    \epsilon^2 \Omega^{(6)}_{\rm near} = \Omega^{(4)}_{\rm mid} = \frac{1}{M} \omega_3(\theta), \qquad \epsilon^2 I^{(6)}_{\rm near} = I^{(4)}_{\rm mid} = \frac{\psi_0}{M} i_3(\theta).
\end{align}
The Znajek condition at the horizon \eqref{Zhor} now says 
\begin{align}\label{i3}
i_3(\theta) = 2\pi \sin^2 \theta \omega_3(\theta).
\end{align}

\subsubsection{Far limit}

At third order in the far limit, Eq.~\eqref{cq} requires
\begin{align}
    \Omega^{(3)}_{\rm far} & = \frac{a}{M^2} \omega_3(\theta) \\ I^{(3)}_{\rm far} & =  \frac{a}{M^2} \left( \frac{-\pi}{2} \psi^{(3)}_{\rm far} \cos \theta + \psi_0 i_3(\theta) \right), \label{I3far}
\end{align}
where we have also matched to the mid limit using \eqref{Iomega2mid}.  (The first term in \eqref{I3far} matches to the first term in \eqref{I3mid} recalling \eqref{nearmatch2} and \eqref{newthing} as well as $\bar{x}=\epsilon x$.)  The stream equation now has a source term,
\begin{align}\label{fareq3}
    L_{\rm far}[\psi_{\rm far}^{(3)}] -  \frac{\psi_0}{16\pi \sin \theta} \pd_\theta \left(\sin^2 \theta g_3(\theta) \right) = s^{(3)}_{\rm far},
\end{align}
with
\begin{align}\label{g3}
     g_3(\theta) &= i_3(\theta) + 2\pi \sin^2 \theta \omega_3(\theta) \\
     & = 4\pi \sin^2 \theta \omega_3(\theta),
\end{align}
(using \eqref{i3} in the second step) and
\begin{align}
    s^{(3)}_{\rm far} & =- \frac{\psi_0 \sin ^2 \theta   \cos  \theta}{\bar{x}^3}.
\end{align}
The boundary condition at small $\bar{x}$ is provided by a match to the mid expansion, which has already been determined to be [we copy \eqref{newthing} here],
\begin{align}\label{bc0}
    \psi^{(3)}_{\rm far} & \sim \frac{\psi_0}{4\bar{x}}\Theta_2(\theta), \qquad \bar{x} \to 0.
\end{align}
The boundary condition at large $\bar{x}$ is that we preserve assumption 2,
\begin{align}\label{bcinf}
    \psi^{(3)}_{\rm far} \sim \textrm{finite}, \qquad \bar{x} \to \infty.
\end{align}
We again have an equation with a single light surface and a single free function $\omega_3(\theta)$, so we expect a unique solution for $\psi_{\rm far}^{(3)}(r,\theta)$ and $\omega_3(\theta)$.  Our numerical analysis below will bear this out.

Before proceeding to the numerical solution, it is useful to note that the Znajek condition at infinity \eqref{Zinf} gives
\begin{align}
    i_3(\theta) + 2\pi  \omega_3(\theta) \sin^2 \theta =  \frac{\pi}{4\psi_0} \left( 2 \cos \theta \psi_\infty^{(3)} - \sin \theta \pd_\theta \psi^{(3)}_{\infty} \right),
\end{align}
where 
\begin{align}\label{psiinfdef}
    \psi^{(3)}_{\infty}(\theta) = \lim_{\bar{x} \to \infty} \psi^{(3)}_{\rm far}.
\end{align}
This can also be derived directly from Eq.~\eqref{fareq3} at large $r$.  Using \eqref{i3} relates $\omega_3(\theta)$ to $\psi^{(3)}_{\infty}(\theta)$ as
\begin{align}\label{W}
    \omega_3(\theta) =  \frac{1}{16 \psi_0 \sin^2 \theta} \left( 2 \cos \theta \psi_\infty^{(3)} - \sin \theta \pd_\theta \psi^{(3)}_{\infty} \right).
\end{align}

Eq.~\eqref{W} allows us to derive analytically that the correction to the total power vanishes at third relative order.  From Eq.~\eqref{dP} in the far limit as $r \to \infty$, we have 
\begin{align}
    P^{(3)}_{\rm far}=-\int_0^\pi \left[I^{(0)}_{\rm far} \Omega^{(0)}_{\rm far}\partial_\theta\psi_{\rm far}^{(3)}+I^{(0)}_{\rm far}\Omega^{(3)}_{\rm far}\partial_\theta\psi_{\rm far}^{(0)}+I^{(3)}_{\rm far}\Omega^{(0)}_{\rm far}\partial_\theta\psi_{\rm far}^{(0)}\right]d\theta.
\end{align}
This integral may be done at any radius $r$.  Each term contributes a non-zero result, but letting $r \to \infty$ and using \eqref{W} shows that the total result vanishes, 
\begin{align}\label{P3}
    P^{(3)}_{\rm far}=2\pi\psi_0\left(\frac{a}{8M^2}\right)^2 \int_0^\pi   \partial_\theta(\sin^2\theta\psi_{\infty}^{(3)})d\theta = 0.
\end{align}
Note that the flux vanishes separately in each hemisphere provided that $\psi^{(3)}$ is odd under $\theta \to \pi-\theta$.  It is expected on general grounds that $\pd_\theta \psi$ is even for the monopole solution, since it is even at leading order and nothing in the problem breaks that symmetry.  Since the corrections $\psi^{(n)}$ must vanish at both poles, there is no freedom to add a constant and it follows that $\psi^{(n)}$ should be odd.  That $\psi^{(3)}_{\rm far}$ should be odd can also be seen from structure of the equation \eqref{fareq3}, since the source $s^{(3)}_{\rm far}$ is odd, while the operator $L_{\rm far}$ preserves partiy.

\begin{figure}
\centering
\includegraphics[width=3.2in]{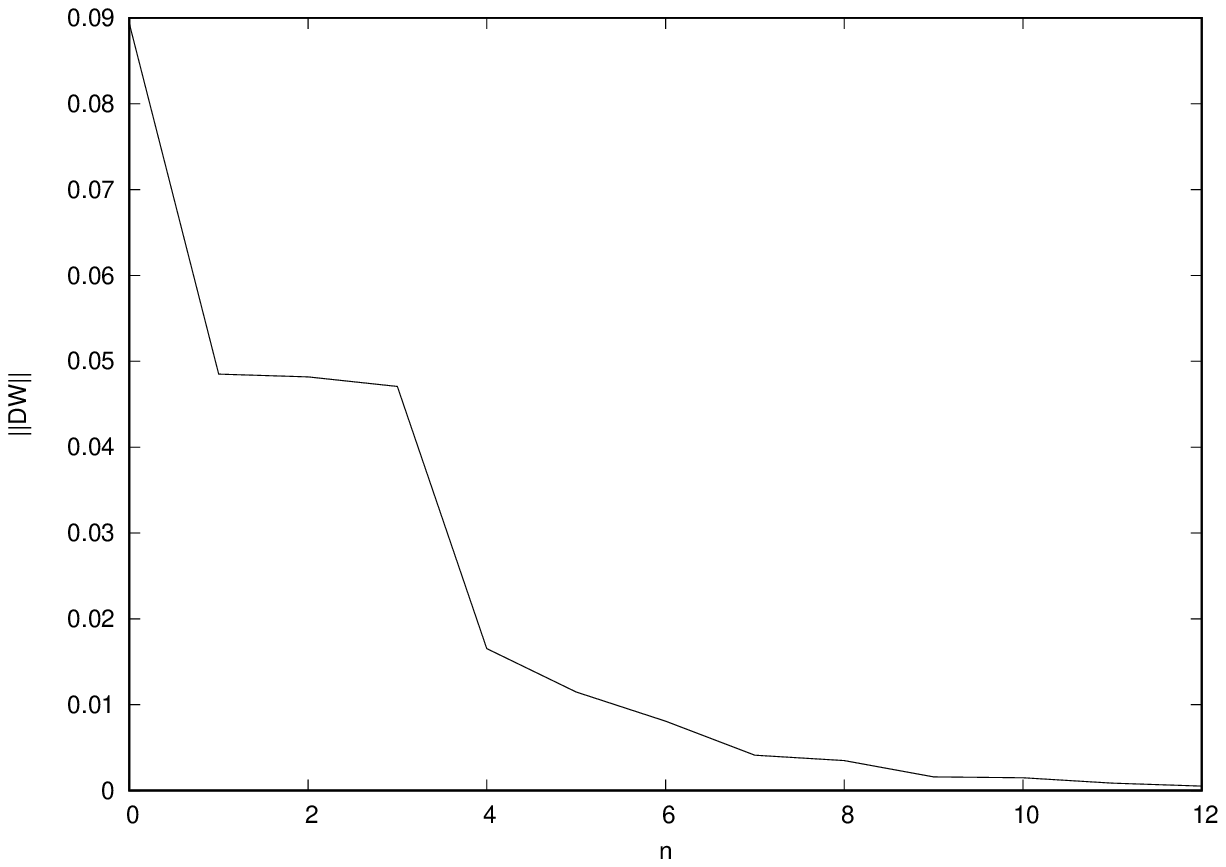} \includegraphics[width=3.2in]{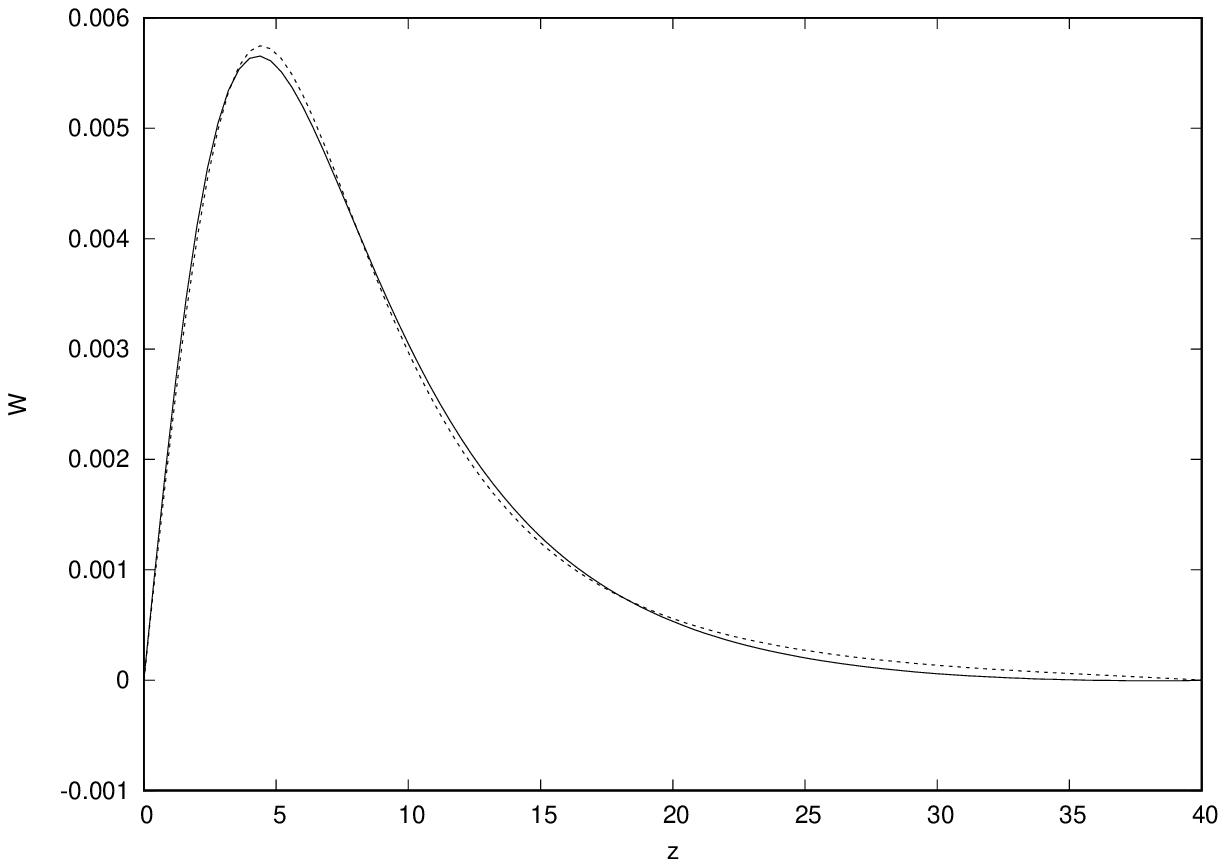}
\caption{Illustration of the parameter optimization.  On the left, we show the light surface mismatch $||\Delta W||$ for the initial guess ($c_2=c_4=c_6=0$), and at each subsequent one dimensional parameter optimization. On the right, we show the two values of $W$ on the light surface as a function of $z \equiv {\bar x} \cos \theta$, for the final iteration where the parameter values are given in Eq.~\eqref{cn}.
}
\label{fig:convergence}
\end{figure}

\begin{figure}
\centering
\includegraphics[width=4.5in]{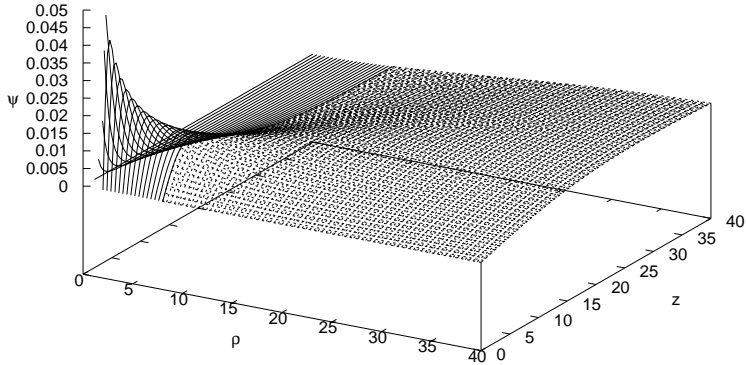}
\caption{The numerical solution for $\psi^{(3)} _{\rm far}/\psi_0$ as a function of $\rho \equiv {\bar x} \sin \theta$ and $z \equiv {\bar x} \cos \theta$.}
\label{psi}
\end{figure}

We now describe our numerical method and present the results; further details are given in App.~\ref{numerics}.  We regard $\psi^{(3)}_\infty(\theta)$ as the free function, with $\omega_3(\theta)$ given by Eq.~\eqref{W}.  We parameterize $\psi^{(3)}_\infty(\theta)$ using the $\Theta_\ell(\theta)$ described in App.~\ref{sec:eigen},
\begin{align}
    \psi^{(3)}_{\infty} = \psi_0 \sum_{\ell=2,4,6,\dots} c_\ell \Theta_\ell(\theta),
\end{align}
where we include only even $\ell$ since $\psi^{(3)}$ should be odd (see discussion below Eq.~\eqref{P3}).  For each choice of $\psi^{(3)}_\infty(\theta)$, we consider the quantity $W(r,\theta) \equiv {\psi ^{(3)} _{\rm far}} - {\psi^{(3)} _\infty}$ and solve Eq.~(\ref{fareq3}) for $W$ separately inside and outside the light surface.  These two solutions generally disagree on the light surface, and we measure the amount of this disagreement by $||\Delta W||$, the $L^2$ norm on the light surface of the difference between the two values of $W$.  In order to obtain a global solution of Eq.~\eqref{fareq3}, we need $||\Delta W||$ to vanish, so we minimize $||\Delta W||$ as a function of the $c_\ell$.  We find that the inferred value of $\psi_\infty^{(3)}$ settles to a fixed function as we increase the number of $c_\ell$ we consider, demonstrating the existence of a solution and providing support for its uniqueness.  In practice, a good fit is provided by including three terms,
\begin{align}\label{psi3inf_3terms}
{\psi ^{(3)}_\infty}(\theta) \approx \psi_0\left( c_2  {\Theta_2}(\theta) + c_4 {\Theta_4}(\theta) + c_6 {\Theta_6}(\theta) \right).
\end{align}
Fig.~\ref{fig:convergence} left shows the behavior of $||\Delta W||$ as a function of the iteration number in parameter optimization, culminating in an optimal set of parameters for which $||\Delta W||$ is very small.  The result is 
\begin{align}\label{cn}
  c_2 =  0.0218, \qquad 
  c_4 = 0.00271, \qquad 
  c_6 = - 0.000316.
\end{align}
Fig.~\ref{fig:convergence} right shows the two values of $W$ on the light surface.  Note the excellent agreement between them.   Given the definition of $W$ and the boundary conditions used, the agreement of the two values of $W$ leads to agreement of the two values of $\psi$ and the gradient of $\psi$.  Thus we have found a numerical global solution of Eq.~\eqref{fareq3} for $\psi ^{(3)} _{\rm far}$.  This solution is shown in Fig.~\ref{psi}.  

The form of $\psi^{(3)}_\infty(\theta)$ determines the function $\omega_3(\theta)$ via Eq.~\eqref{W}.  Plugging Eq.~\eqref{psi3inf_3terms} into Eq.~\eqref{W} gives
\begin{align}\label{omega3_full}
    \omega_3(\theta) & \approx \frac{35c_2-14c_4+8c_6}{560}\Theta_1(\theta)+\frac{7c_4-4c_6}{80}\Theta_3(\theta)+\frac{11c_6}{112}\Theta_5(\theta),\\
    & = .0013 \Theta_1(\theta)+.00025\Theta_3(\theta)-.000031\Theta_5(\theta),
\end{align}
where we use \eqref{cn} in the second line.

\section{Comparison with Previous work}\label{sec:previous}

 Using what we call the mid expansion, Blandford and Znajek \cite{blandford-znajek1977} obtained the leading order solution and energy extraction rate.  Their strategy was to demand that the solution obey the same relationship between $I$ and $\Omega$ as a corresponding flat spacetime solution (in this case the Michel monopole).  Our approach clarifies the meaning of this step: it is imposed by smooth matching to the far limit.  Blandford and Znajek also found the first non-vanishing correction to the flux function, which we denote $\psi^{(2)}_{\rm mid}$.

Tanabe and Nagataki \cite{tanabe-nagataki2008} later found a solution for $I^{(3)}_{\rm mid}$, $\Omega^{(3)}_{\rm mid}$, and $\psi^{(4)}_{\rm mid}$ which contained two undetermined parameters.  They noted that no value of these parameters could prevent  $\psi^{(4)}_{\rm mid}$ from diverging at large $r$, and concluded that a better understanding of boundary conditions at infinity was necessary.  Interestingly, they were able to obtain a unique energy extraction rate from their partial solution, apparently due to a coincidental cancelation of terms involving the undetermined parameters.  Our work settles the issue of boundary conditions---one must match to the far expansion and then impose finiteness of $\psi$ at infinity (equivalently \eqref{Zinf})---and confirms the subleading energy extraction rate first calculated by Tanabe and Nagataki.

 Pan and Yu \cite{pan-yu2015} later found the full expression for $I^{(3)}_{\rm mid}$ and $\Omega^{(3)}_{\rm mid}$ by imposing the condition that the fourth-order source term for the stream equation must vanish at large $r$ (Eqs.~(29) and (30) therein).  While this happens gives the correct answer because (as we show) this source does happen to fall off (Eqs.~\eqref{s4mid}, \eqref{f2large}, and \eqref{f4large} below), we see no justification for imposing the falloff condition (and none is given in Ref.~\cite{pan-yu2015}).  Furthermore, to derive results at any given order in perturbation theory, it should never be necessary to appeal to properties at higher order.  Indeed, our work shows that $I^{(3)}_{\rm mid}$ and $\Omega^{(3)}_{\rm mid}$ may be straightforwardly derived using only the relative order to which they belong (as well as lower orders).  Our work shows that, despite the flawed derivation in Ref.~\cite{pan-yu2015}, the results are in fact correct.
 
 Pan and Yu later applied their method at higher order in Ref.~\cite{pan-yu2015b}.  This paper contains unjustified assumptions and  erroneous claims.  The unjustified falloff condition  (``convergence condition'') is used at all orders.  Expressed in our notation, the authors claim that $\psi^{(4)}_{\rm mid}$ vanishes at large $r$ (Eq.~(18) therein).  Three independent analyses (Ref.~\cite{tanabe-nagataki2008}, Ref.~\cite{orselli-etal2018} and Eqs.~\eqref{Rhat42} and \eqref{Rhat44} below) have shown that this claim is false: $\psi^{(4)}_{\rm mid}$ in fact diverges at large $r$.  The authors then make the same claim at all orders, asserting (in our notation) that $\psi(r \to \infty)=\psi_0(1-\cos \theta)$ \textit{exactly} (see discussion above Eq.~(30) therein), or equivalently that all perturbations to $\psi$ vanish at infinity.  This claim is false both in the mid limit they consider (as described above for $\psi^{(4)}_{\rm mid}$) and in the far limit that is necessary to actually access the asymptotic region, as we show by finding a non-zero value for $\psi^{(3)}_{\infty}$ in Eq.~\eqref{psi3inf_3terms}.

The pertubative monopole solution was revisited more recently in Ref.~\cite{orselli-etal2018}.  The authors also considered a perturbative approach they termed ``matched asymptotic expansions'', and concluded that this approach fails.  In terms of our notation, they assume a jointly smooth expansion in ($\epsilon$, $M/r$).  We would call this an assumption of overlapping regular expansions (large $r$ and small $\epsilon$), rather than the method of matched asymptotic expansions, because there is no ``mixing'' between the perturbation parameter $\epsilon$ and the radius $r$.  The analysis of \cite{orselli-etal2018} shows that this assumption of overlapping regular expansions is inconsistent, indicating the need for a true matched asymptotic expansions approach like ours.  (Ref.~\cite{orselli-etal2019} further demonstrates the need for matched asymptotic expansions.) Our work shows that the monopole flux function is jointly $C^3$ in ($\epsilon x$, $1/x$), where $x=r/M$,\footnote{Let $a=\epsilon x$ and $b=1/x$.  The mid limit $\epsilon \to 0$ fix $x$ is equivalent to $a\to0$ fix $b$, and the far limit $\epsilon \to 0$ fix $\epsilon x$ is equivalent to $b \to 0$ fix $a$.  The behavior we find in the overlap region corresponds to Taylor expansion of $\psi$ in $a$ and $b$, demonstrating the joint smoothness.} with similar statements for the other quantities.  However, at next order logarithms will appear, and ultimately the near limit may become important, so we make no claims about the behavior of the solution family to higher order.  See Sec.~II of Ref.~\cite{gralla-wald2008} for further discussion of the relationship between matched asymptotic expansions and jointly smooth behavior in ``mixed'' variables.

\section*{Acknowledgements} We thank Ioannis Contopoulos for helpful correspondence. This work was supported in part by NSF grants PHY-1506027, PHY-1752809, and PHY-1806219. G.C. is research associate of the F.R.S.-FNRS and he acknowledges support from the IISN convention 4.4503.15, the COST Action GWverse CA16104 and FNRS conventions R.M005.19 and J.0036.20. J.A. is partly supported by the Netherlands Organization for Scientific Research (NWO) and by the Dutch Institute for Emergent Phenomena (DIEP) cluster at the University of Amsterdam.

\appendix

\section{Mid limit at fourth order}\label{sec:L4}

We now discuss the mid limit at one order higher than considered in the text.  This allows us to connect with previous work and also speculate about the behavior of the solution at higher order.  First we note that \textit{if} there is a smooth expansion through fourth order in $\epsilon$ in the mid limit, the flux function will satisfy
\begin{align}\label{eq4mid}
    L_{\rm mid}[\psi^{(4)}_{\rm mid}]=s^{(4)}_{\rm mid},
\end{align}
where the source term is
\begin{align}\label{s4mid}
    s^{(4)}_{\rm mid}=\psi_0 f_2(x)\Theta_2(\theta)+\psi_0 f_4(x)\Theta_4(\theta)
\end{align}
with
\begin{align}
    f_2(x)&=\frac{1}{112(x-2)x^6}(192(x-2)x^4 \hat{R}(x)^2\nonumber \\
    &-2x^2\hat{R}(x)(-704+480x-64x^2+3x^5+32(x-2)x^2\hat{R}^{'}(x)+16(x-2)^2x^3\hat{R}^{''}(x))\nonumber\\
    &+48\hat{R}(2)x^4+2(-288+112x+72x^2-16x^4+x^5)\nonumber\\
    &+2x^2(-384+192x-32x^2+16x^3-2x^5+x^6)\hat{R}^{'}(x)\nonumber\\
    &+x^3(384-288x+112x^2-32x^3-2x^5+x^6)\hat{R}^{''}(x))
\end{align}
and
\begin{align}
    f_4(x)&=\frac{3}{448(x-2)x^6}(-1152(x-2)x^4\hat{R}^2(x)\nonumber\\
    &-2x^2\hat{R}(x)(3(512-64x-96x^2+x^5)-192(x-2)x^2\hat{R}^{'}(x)-96(x-2)^2x^3\hat{R}^{''}(x))\nonumber\\
    &+48\hat{R}(2)x^4+2(384-96x^2-2x^4+x^5)+2(x-2)x^2(-256+16x^2+x^5)\hat{R}^{'}(x)\nonumber\\
    &+(x-2)x^3(256-64x-32x^2+x^5)\hat{R}^{''}(x)).
\end{align}
The function $\hat{R}(x)$ was given in Eq.~\eqref{Rhat} above.  At large $x$ we have 
\begin{align}\label{f2large}
    f_2(x) & \sim -\frac{3}{224x}+O\left(\frac{\log x}{x^2}\right), \quad    x \to \infty\\
    f_4(x) & \sim -\frac{9}{896x}+O\left(\frac{\log x}{x^2}\right),  \quad   x \to \infty.\label{f4large}
\end{align}
Analogously to \eqref{psi2gen}, we write the general solution as 
\begin{align}\label{psi4gen}
    \psi^{(4)}_{\rm mid}=\psi_0 \hat{R}_2(x) \Theta_2(\theta)+\psi_0 \hat{R}_4(x) \Theta_4(\theta) +\sum_{\ell=1}^\infty \left( B^<_\ell R^<_\ell(x) + B^>_\ell R^>_\ell(x) \right) \Theta_\ell(\theta),
\end{align}
where $\hat{R}_2$ and $\hat{R}_4$ are particular solutions to the $\ell=2$ and $\ell=4$ equations,
\begin{align}
    \hat{R}_2(x)&=-2R^{<}_2(x) \int R^{>}_2(x)f_2(x)dx + 2R^{>}_2(x) \int R^{<}_2(x)f_2(x)dx,\label{Rhat42}\\
    \hat{R}_4(x)&=-8R^{<}_4(x) \int R^{>}_4(x)f_4(x)dx + 8R^{>}_4(x) \int R^{<}_4(x)f_4(x)dx.\label{Rhat44}
\end{align}
Expanding the solutions near infinity gives 
\begin{align}
    \hat{R}_2(x) &\sim C_2^a\left(x^3-\frac{3}{2}x^2\right)+\frac{x}{448}+\frac{227-60\log(2)+60\log(x)}{100800}+O\left(\frac{\log x}{x}\right), \label{pony1} \\
    \hat{R}_4(x) &\sim C_4^a\left(\frac{x^5}{6}-\frac{5x^4}{8}+\frac{5x^3}{84}-\frac{5x^2}{21}\right)+\frac{9x}{17920}-\frac{3(-121+40\log(2)-40\log(x))}{896000}+O\left(\frac{\log x}{x}\right),\label{pony2}
\end{align}
where $C_2^a$ and $C_4^a$ are constants of integration that are degenerate with $B_2^<$ and $B_4^<$ in Eq.~\eqref{psi4gen}.  Even if we set these constants to zero, the solution still blows up like $x$, as first observed in Ref.~\cite{tanabe-nagataki2008}.  In a mid-only perturbation approach the blowup indicates an inconsistency.  Our approach has resolved the inconsistency by including the far limit and noting that there is a contribution at third relative order, i.e. one order \textit{lower} than the presumed inconsistency.  Note, however, that we infer from the third order far equation \eqref{fareq3} that
\begin{align}
 \psi_{\rm far}^{(3)} \stackrel{\bar x \to 0}{\sim} \frac{\psi_0 \Theta_2(\theta)}{4} \frac{1}{\bar x} +\psi_0 \left(\frac{\Theta_2(\theta)}{448}+\frac{9\Theta_4(\theta)}{17920}\right)\bar x+O(\bar x^2). 
\end{align}
We already discussed that the leading $1/\bar x$ term matches with $\psi^{(2)}_{\rm mid}$ in the matching region between the middle and far region. The subleading $\bar x$ term now matches with the large radius behavior of $\psi^{(4)}_{\rm mid}$ derived in \eqref{pony1}-\eqref{pony2} after setting $C_2^a=C_4^a=0$.

In this work we do not go beyond the third relative order.  However, we can anticipate already that the next order will involve logarithms, such that the assumption of a smooth expansion that underlies Eq.~\eqref{eq4mid} is unlikely to be correct.  The reason is that the solutions \eqref{pony1} and \eqref{pony2} contain $\log x$ terms which contribute terms of order $\epsilon^4 \log x=\epsilon^4 \log(\bar{x}/\epsilon)$ to the flux function, indicating the need for an $\epsilon^4 \log \epsilon$ term in the far expansion.  In cases like this one typically circles back and includes logarithms in the next-order ansatz for all expansions in order to consistently derive an $\epsilon^4 \log \epsilon$ relative correction.  We leave such analysis for future work. 

\section{Eigenfunctions of the mid operator}\label{sec:eigen}
 
 An important operator that appears in our analysis is
 \begin{align}
    L_{\rm mid}={\partial}_x [(1-\frac{2}{x}){\partial}_x]+\frac{\sin\theta}{x^2}{\partial}_\theta[\frac{1}{\sin\theta}{\partial}_\theta].
\end{align}
This operator separates into
\begin{align}
    L^{\theta}_\ell&={\partial}_\theta(\frac{1}{\sin\theta}{\partial}_\theta)+\frac{\ell(\ell+1)}{\sin\theta} \label{Ltheta} \\
    L^{x}_\ell&={\partial}_x [(1-\frac{2}{x}){\partial}_x]-\frac{\ell(\ell+1)}{x^2}.\label{Lx}
\end{align}
These operators are analyzed in  Ref.~\cite{gralla-lupsasca-rodriguez2015}, and we present the needed results here.

\subsection{Angular functions} 
Solutions $L^\theta_\ell[\Theta(\theta)]=0$ that vanish quadratically at both poles only occur for integers $\ell \geq 1$, and are given by (with $k>0$ a positive integer)
\begin{align}
    \Theta_{2k-1}(\theta)&= \,_2F_1[-k,k-\frac{1}{2};\frac{1}{2};\cos^2\theta]\\
    \Theta_{2k}(\theta)&=\,_2F_1[-k,k+\frac{1}{2};\frac{3}{2};\cos^2\theta]\cos\theta.
\end{align}
The first several eigenfunctions are
\begin{align}
    \Theta_1(\theta) & = \sin^2 \theta \\
    \Theta_2(\theta) & = \cos \theta \sin^2 \theta \\
    \Theta_3(\theta) & = (1-5\cos^2 \theta) \sin^2 \theta \\
    \Theta_4(\theta) & = (1-\tfrac{7}{3} \cos^2\theta)\cos \theta \sin^2 \theta.
\end{align}
These $\Theta_\ell$ are orthogonal with weight $\csc \theta$.  We presume they are also complete for functions vanishing at both poles, but the differential operator \eqref{Ltheta} is not sufficiently regular to apply the standard Strum-Liouville theorems.

\subsection{Radial functions}
For $\ell>0$, the general solution $R_\ell$ to the homogeneous radial equation $L^x_\ell[R_\ell]=0$ is a linear combination of
 \begin{align}
     R^{<}_\ell(x)&=\frac{x^2}{2}\frac{{\Gamma (\ell+2)}^2}{\Gamma (2\ell+1)}\, _2F_1[\ell+2,1-\ell;3;\frac{x}{2}]\\
     R^{>}_\ell(x)&=-\frac{2}{\sqrt{\pi}}\left(\frac{x}{4}\right)^{-\ell}\left\{\, _2F_1[\ell+2,\ell;1;1-\frac{2}{x}]\log\left(1-\frac{2}{x}\right)+P_\ell\left(\frac{x}{2}\right)\right\},
 \end{align}
 where the polynomials $P_\ell$ are defined recursively by 
 \begin{align}
     P_1(x)&=x^2+\frac{x}{2}\\
     P_2(x)&=4x^4-x^3-\frac{x^2}{6}\\
     P_\ell(x)&=\frac{(2\ell-1)[\ell(\ell-1)(2x-1)-1]x P_{\ell-1}(x)-\ell^2(\ell-2)x^2 P_{\ell-2}(x)}{(\ell+1)(\ell-1)^2}.
 \end{align}
 The normalization has been chosen so that
 \begin{align}
     R^<_\ell(x) &\sim x^{\ell+1}, \qquad x \to \infty \\
     R^{>}_\ell(x) &\sim x^{-\ell}, \ \qquad x \to \infty.
 \end{align}

\section{Numerical Methods}\label{numerics}

We now present the numerical methods used to solve Eq.~\eqref{fareq3} for $\psi ^{(3)} _{\rm far}$.  We will find it helpful to introduce the quantity $W$ defined by
\begin{equation}
W \equiv {\psi ^{(3)} _{\rm far}} - {\psi ^{(3)} _\infty}.
\end{equation}
Then Eq.~\eqref{fareq3} becomes
\begin{equation}
{L_{\rm far}}[W] = - {\frac {{\psi _0} {\sin ^2} \theta \cos \theta} {{\bar x}^3}} \; - \; {\frac {\sin \theta}
{{\bar x}^2}} {\frac d {d \theta}} \left ( {\frac 1 {\sin \theta}} {\frac {d {\psi ^{(3)} _\infty}} {d \theta}}
\right ).
\label{Weqn1}
\end{equation}
Here we have used Eqs.~\eqref{g3} and \eqref{W} to express ${g_3}(\theta)$ in terms of ${\psi ^{(3)} _\infty}(\theta)$.  

We will also find it helpful to introduce cylindrical coordinates $(\rho,z)$ given by 
\begin{equation}
\rho \equiv {\bar x} \sin \theta , \; \; \; z \equiv {\bar x} \cos \theta .
\end{equation}
Then Eq.~\eqref{Weqn1} becomes
\begin{eqnarray}
\left ( 1 - {\frac {\rho ^2} {64}} \right ) \left [ {\frac {{\partial ^2} W} {\partial {\rho ^2}}} \; + \; 
{\frac {{\partial ^2} W} {\partial {z^2}}} \right ] - \left ( {\frac 1 \rho} + {\frac \rho {64}} \right ) 
{\frac {\partial W} {\partial \rho}} + {\frac 1 {32}} (2 - 3 {\sin ^2}\theta) W 
\nonumber
\\
= 
- {\frac {{\psi _0} {\sin ^2} \theta \cos \theta} {{\bar x}^3}} \; - \; {\frac {\sin \theta}
{{\bar x}^2}} {\frac d {d \theta}} \left ( {\frac 1 {\sin \theta}} {\frac {d {\psi ^{(3)} _\infty}} {d \theta}}
\right ).
\label{Weqn2}
\end{eqnarray}
Here $\bar x$ and $\theta$ are to be thought of as functions of $\rho$ and $z$ given by ${\bar x} = {\sqrt {{\rho ^2} + {z^2}}}$ and $\theta = {\tan ^{-1}}(\rho /z)$. 

Eq.~\eqref{Weqn2} is a singular elliptic equation.  That is, this equation is elliptic except at $\rho =8$.  Furthermore, it is clear that at $\rho=8$ any nonsingular solution must satisfy
\begin{equation}
{\frac {\partial W} {\partial \rho}} - {\frac 1 8} (2 - 3 {\sin ^2}\theta) W = 
 {\frac {4{\psi _0} {\sin ^2} \theta \cos \theta} {{\bar x}^3}} \; + \; {\frac {4\sin \theta}
{{\bar x}^2}} {\frac d {d \theta}} \left ( {\frac 1 {\sin \theta}} {\frac {d {\psi ^{(3)} _\infty}} {d \theta}}
\right ).
\label{boundary}
\end{equation} 

Therefore, one should think of Eq.~\eqref{Weqn2} as {\emph {two}} singular elliptic equations: one defined on an inner domain where $0 \le \rho \le 8$ and one defined on an outer domain where $ 8 \le \rho < \infty$.  In each case $\rho =8$ is a boundary of the domain on which one imposes the boundary condition of Eq.~\eqref{boundary}).  We know of no general theorems on singular elliptic equations that would allow us to deduce existence or uniqueness for each of the two singular elliptic problems.  Nonetheless, we can apply the standard numerical method of relaxation to the problems and see what happens.  

Relaxation is an iterative method that works as follows: The function is represented as a set of values at regularly spaced grid points.  Eq.~\eqref{Weqn2} then becomes an expression for the value of the function at each grid point in terms of the values at each of its nearest neighbors.  This expression is used to give the value of the function at the next iteration.  The method computes a residual that is a measure of how badly the function at the current iteration fails to satisfy the finite difference version of the differential equation.  Once the residual falls below some preset tolerance, the program declares that it has found a solution and the iteration terminates.  If after a preset number of iterations, the specified tolerance is not reached, the program declares failure to find a solution and the program halts.  The iteration actually reaching the specified tolerance should be taken as {\it prima facie} evidence both that the equation has a solution and that the numerical method has found it.

We now want to be more precise about the boundary conditions for each of the domains.  Though $W$ is defined on an infinite size domain, computer grids are finite.  We therefore introduce quantities $\rho_{\rm max}$ and $z_{\rm max}$ as the maximum values of $\rho$ and $z$ respectively.  By the definition of $W$ it follows that $W \to 0$ as $\rho \to \infty$ or $z \to \infty$.  We will implement this condition numerically by requiring that $W=0$ at $\rho = {\rho_{\rm max}}$ and at $z = {z_{\rm max}}$.  From Eq.~\eqref{bc0} it follows that $W$ is singular at small $\bar x$.  So we introduce the quantity ${\bar x}_{\rm min}$ as the minimum allowed value of $\bar x$ in the domain, and we impose the condition that at that boundary 
\begin{equation}
W = {\frac {{\psi_0} {\sin ^2}\theta \cos \theta} {4 {\bar x}}} - {\psi ^{(3)} _\infty}
\label{Wsmall}
\end{equation}   
Finally, the angular dependence of $\psi ^{(3)}$ is such that it vanishes at $\theta =0 $ and at $\theta = \pi/2$, from which we conclude that $W$ vanishes at $\rho=0$ and at $z=0$.  To summarize: the inner domain has boundaries at $\rho=0, \, z=0$ and $z= {z_{\rm max}}$ at which $W$ vanishes, a boundary at ${\bar x} = {{\bar x}_{\rm min}}$ at which Eq.~\eqref{Wsmall} is imposed, and a boundary at $\rho=8$ at which Eq.~\ref{boundary} is imposed.  The outer domain has boundaries at $z=0, \, z = {z_{\rm max}}$ and $\rho = {\rho _{\rm max}}$ at which $W$ vanishes, and a boundary at $\rho = 8$ at which Eq.~\eqref{boundary} is imposed.

It is clear that this numerical method is making multiple approximations. Thus we expect to obtain a solution of Eq.~\eqref{Weqn1} only in the simultaneous limit in which grid spacing goes to zero, ${\bar x}_{\rm min} \to 0$ and $\rho_{\rm max}$ and $z_{\rm max}$ go to infinity.

Having solved Eq.~\eqref{Weqn2} on both the inner domain and the outer domain, subject to the appropriate boundary conditions, we still do not have a global solution to Eq.~\eqref{Weqn2}.  The reason is that in general the value of $W$ at $\rho=8$ for the inner domain (which we will call $W_1$) will not be the same as the value of $W$ at $\rho =8$ for the outer domain (which we will call $W_2$).  However, we have at our disposal the function ${\psi^{(3)}_\infty}$ which we will choose in an attempt to make $W_1$ equal to $W_2$.  More precisely, we define $||\Delta W||$ to be the $L^2$ norm of ${W_2}-{W_1}$, and we pick a parametrized space of possible ${\psi ^{(3)} _\infty}$.  We then find the value of the parameters that lead to the smallest $||\Delta W||$.  As usual with such numerical methods, if we can get $||\Delta W||$ sufficiently small, we declare that we have found an approximate solution.  (And if we can't, then we either declare failure to find a solution, or we look for a better parameter space).  We will choose the following parameter space
\begin{equation}
{\psi ^{(3)} _\infty}(\theta) = {c_2} {\Theta_2}(\theta) + {c_4} {\Theta _4}(\theta) + {c_6} {\Theta _6} (\theta)
\end{equation}    
Where $c_2, \, c_4$ and $c_6$ are the parameters.  We find the minimum by applying the line minimization method given in Numerical Recipes \cite{numerical-recipes}.  That is, we start out with guesses for $c_2 , \, c_4$ and $c_6$.  Then keeping $c_4$ and $c_6$ fixed we use the Numerical Recipes one dimensional search to find the value of $c_2$ that minimizes $||\Delta W||$.  Then with that value of $c_2$, we pick the $c_4$ that gives the smallest $||\Delta W||$.  Then on to $c_6$, then back to $c_2$ and so on until $||\Delta W||$ isn't getting any smaller, at which point we can declare victory if $||\Delta W||$ is sufficiently small.

\bibliographystyle{utphys}
\bibliography{BZmethod}

\providecommand{\href}[2]{#2}\begingroup\raggedright\begin{thebibliography}{10}

\bibitem{blandford-znajek1977}
R.~D. {Blandford} and R.~L. {Znajek}, ``{Electromagnetic extraction of energy
  from Kerr black holes},'' {\em \mnras} {\bfseries 179} (May, 1977) 433--456.

\bibitem{EHTcodeComparison}
O.~{Porth}, K.~{Chatterjee}, R.~{Narayan}, C.~F. {Gammie}, Y.~{Mizuno},
  P.~{Anninos}, J.~G. {Baker}, M.~{Bugli}, C.-k. {Chan}, J.~{Davelaar}, L.~{Del
  Zanna}, Z.~B. {Etienne}, P.~C. {Fragile}, B.~J. {Kelly}, M.~{Liska},
  S.~{Markoff}, J.~C. {McKinney}, B.~{Mishra}, S.~C. {Noble}, H.~{Olivares},
  B.~{Prather}, L.~{Rezzolla}, B.~R. {Ryan}, J.~M. {Stone}, N.~{Tomei}, C.~J.
  {White}, Z.~{Younsi}, K.~{Akiyama}, A.~{Alberdi}, W.~{Alef}, K.~{Asada},
  R.~{Azulay}, A.-K. {Baczko}, D.~{Ball}, M.~{Balokovi{\'c}}, J.~{Barrett},
  D.~{Bintley}, L.~{Blackburn}, W.~{Boland}, K.~L. {Bouman}, G.~C. {Bower},
  M.~{Bremer}, C.~D. {Brinkerink}, R.~{Brissenden}, S.~{Britzen}, A.~E.
  {Broderick}, D.~{Broguiere}, T.~{Bronzwaer}, D.-Y. {Byun}, J.~E. {Carlstrom},
  A.~{Chael}, S.~{Chatterjee}, M.-T. {Chen}, Y.~{Chen}, I.~{Cho},
  P.~{Christian}, J.~E. {Conway}, J.~M. {Cordes}, {Geoffrey}, B.~{Crew},
  Y.~{Cui}, M.~{De Laurentis}, R.~{Deane}, J.~{Dempsey}, G.~{Desvignes}, S.~S.
  {Doeleman}, R.~P. {Eatough}, H.~{Falcke}, V.~L. {Fish}, E.~{Fomalont},
  R.~{Fraga-Encinas}, B.~{Freeman}, P.~{Friberg}, C.~M. {Fromm}, J.~L.
  {G{\'o}mez}, P.~{Galison}, R.~{Garc{\'\i}a}, O.~{Gentaz}, B.~{Georgiev},
  C.~{Goddi}, R.~{Gold}, M.~{Gu}, M.~{Gurwell}, K.~{Hada}, M.~H. {Hecht},
  R.~{Hesper}, L.~C. {Ho}, P.~{Ho}, M.~{Honma}, C.-W.~L. {Huang}, L.~{Huang},
  D.~H. {Hughes}, S.~{Ikeda}, M.~{Inoue}, S.~{Issaoun}, D.~J. {James}, B.~T.
  {Jannuzi}, M.~{Janssen}, B.~{Jeter}, W.~{Jiang}, M.~D. {Johnson},
  S.~{Jorstad}, T.~{Jung}, M.~{Karami}, R.~{Karuppusamy}, T.~{Kawashima}, G.~K.
  {Keating}, M.~{Kettenis}, J.-Y. {Kim}, J.~{Kim}, J.~{Kim}, M.~{Kino}, J.~Y.
  {Koay}, {Patrick}, M.~{Koch}, S.~{Koyama}, M.~{Kramer}, C.~{Kramer}, T.~P.
  {Krichbaum}, C.-Y. {Kuo}, T.~R. {Lauer}, S.-S. {Lee}, Y.-R. {Li}, Z.~{Li},
  M.~{Lindqvist}, K.~{Liu}, E.~{Liuzzo}, W.-P. {Lo}, A.~P. {Lobanov},
  L.~{Loinard}, C.~{Lonsdale}, R.-S. {Lu}, N.~R. {MacDonald}, J.~{Mao}, D.~P.
  {Marrone}, A.~P. {Marscher}, I.~{Mart{\'\i}-Vidal}, S.~{Matsushita}, L.~D.
  {Matthews}, L.~{Medeiros}, K.~M. {Menten}, I.~{Mizuno}, J.~M. {Moran},
  K.~{Moriyama}, M.~{Moscibrodzka}, C.~{M{\"u}ller}, H.~{Nagai}, N.~M. {Nagar},
  M.~{Nakamura}, G.~{Narayanan}, I.~{Natarajan}, R.~{Neri}, C.~{Ni},
  A.~{Noutsos}, H.~{Okino}, T.~{Oyama}, F.~{{\"O}zel}, D.~C.~M. {Palumbo},
  N.~{Patel}, U.-L. {Pen}, D.~W. {Pesce}, V.~{Pi{\'e}tu}, R.~{Plambeck},
  A.~{PopStefanija}, J.~A. {Preciado-L{\'o}pez}, D.~{Psaltis}, H.-Y. {Pu},
  V.~{Ramakrishnan}, R.~{Rao}, M.~G. {Rawlings}, A.~W. {Raymond},
  B.~{Ripperda}, F.~{Roelofs}, A.~{Rogers}, E.~{Ros}, M.~{Rose},
  A.~{Roshanineshat}, H.~{Rottmann}, A.~L. {Roy}, C.~{Ruszczyk}, K.~L.~J.
  {Rygl}, S.~{S{\'a}nchez}, D.~{S{\'a}nchez-Arguelles}, M.~{Sasada},
  T.~{Savolainen}, F.~P. {Schloerb}, K.-F. {Schuster}, L.~{Shao}, Z.~{Shen},
  D.~{Small}, B.~W. {Sohn}, J.~{SooHoo}, F.~{Tazaki}, P.~{Tiede}, R.~P.~J.
  {Tilanus}, M.~{Titus}, K.~{Toma}, P.~{Torne}, T.~{Trent}, S.~{Trippe},
  S.~{Tsuda}, I.~{van Bemmel}, H.~J. {van Langevelde}, D.~R. {van Rossum},
  J.~{Wagner}, J.~{Wardle}, J.~{Weintroub}, N.~{Wex}, R.~{Wharton},
  M.~{Wielgus}, G.~N. {Wong}, Q.~{Wu}, K.~{Young}, A.~{Young}, F.~{Yuan}, Y.-F.
  {Yuan}, J.~A. {Zensus}, G.~{Zhao}, S.-S. {Zhao}, Z.~{Zhu}, and {(The Event
  Horizon Telescope Collaboration}, ``{The Event Horizon General Relativistic
  Magnetohydrodynamic Code Comparison Project},''
  \href{http://dx.doi.org/10.3847/1538-4365/ab29fd}{{\em \apjs} {\bfseries 243}
  no.~2, (Aug, 2019) 26}, \href{http://arxiv.org/abs/1904.04923}{{\ttfamily
  arXiv:1904.04923 [astro-ph.HE]}}.

\bibitem{parfrey-philippov-cerutti2019}
K.~{Parfrey}, A.~{Philippov}, and B.~{Cerutti}, ``{First-Principles Plasma
  Simulations of Black-Hole Jet Launching},''
  \href{http://dx.doi.org/10.1103/PhysRevLett.122.035101}{{\em \prl} {\bfseries
  122} no.~3, (Jan, 2019) 035101},
  \href{http://arxiv.org/abs/1810.03613}{{\ttfamily arXiv:1810.03613
  [astro-ph.HE]}}.

\bibitem{komissarov2001}
S.~S. {Komissarov}, ``{Direct numerical simulations of the Blandford-Znajek
  effect},'' \href{http://dx.doi.org/10.1046/j.1365-8711.2001.04863.x}{{\em
  \mnras} {\bfseries 326} (Sept., 2001) L41--L44}.

\bibitem{tchekhovskoy-narayan-mckinney2010}
A.~{Tchekhovskoy}, R.~{Narayan}, and J.~C. {McKinney}, ``{Black Hole Spin and
  The Radio Loud/Quiet Dichotomy of Active Galactic Nuclei},''
  \href{http://dx.doi.org/10.1088/0004-637X/711/1/50}{{\em \apj} {\bfseries
  711} (Mar., 2010) 50--63}, \href{http://arxiv.org/abs/0911.2228}{{\ttfamily
  arXiv:0911.2228 [astro-ph.HE]}}.

\bibitem{nathanail-contopoulos2014}
A.~{Nathanail} and I.~{Contopoulos}, ``{Black Hole Magnetospheres},''
  \href{http://dx.doi.org/10.1088/0004-637X/788/2/186}{{\em \apj} {\bfseries
  788} (June, 2014) 186}, \href{http://arxiv.org/abs/1404.0549}{{\ttfamily
  arXiv:1404.0549 [astro-ph.HE]}}.

\bibitem{tanabe-nagataki2008}
K.~{Tanabe} and S.~{Nagataki}, ``{Extended monopole solution of the
  Blandford-Znajek mechanism: Higher order terms for a Kerr parameter},''
  \href{http://dx.doi.org/10.1103/PhysRevD.78.024004}{{\em \prd} {\bfseries 78}
  no.~2, (July, 2008) 024004}, \href{http://arxiv.org/abs/0802.0908}{{\ttfamily
  arXiv:0802.0908}}.

\bibitem{pan-yu2015}
Z.~{Pan} and C.~{Yu}, ``{Fourth-order split monopole perturbation solutions to
  the Blandford-Znajek mechanism},''
  \href{http://dx.doi.org/10.1103/PhysRevD.91.064067}{{\em \prd} {\bfseries 91}
  no.~6, (Mar., 2015) 064067},
  \href{http://arxiv.org/abs/1503.05248}{{\ttfamily arXiv:1503.05248
  [astro-ph.HE]}}.

\bibitem{pan-yu2015b}
Z.~{Pan} and C.~{Yu}, ``{Analytic Properties of Force-free Jets in the Kerr
  Spacetime - I},'' \href{http://dx.doi.org/10.1088/0004-637X/812/1/57}{{\em
  \apj} {\bfseries 812} (Oct., 2015) 57},
  \href{http://arxiv.org/abs/1504.04864}{{\ttfamily arXiv:1504.04864
  [astro-ph.HE]}}.

\bibitem{orselli-etal2018}
G.~{Grignani}, T.~{Harmark}, and M.~{Orselli}, ``{Existence of the
  Blandford-Znajek monopole for a slowly rotating Kerr black hole},''
  \href{http://dx.doi.org/10.1103/PhysRevD.98.084056}{{\em \prd} {\bfseries 98}
  no.~8, (Oct, 2018) 084056}, \href{http://arxiv.org/abs/1804.05846}{{\ttfamily
  arXiv:1804.05846 [gr-qc]}}.

\bibitem{orselli-etal2019}
G.~{Grignani}, T.~{Harmark}, and M.~{Orselli}, ``{Force-free electrodynamics
  near rotation axis of a Kerr black hole},'' {\em arXiv e-prints} (Aug, 2019)
  arXiv:1908.07227, \href{http://arxiv.org/abs/1908.07227}{{\ttfamily
  arXiv:1908.07227 [gr-qc]}}.

\bibitem{gralla-lupsasca-philippov2016}
S.~E. {Gralla}, A.~{Lupsasca}, and A.~{Philippov}, ``{Pulsar Magnetospheres:
  Beyond the Flat Spacetime Dipole},''
  \href{http://dx.doi.org/10.3847/1538-4357/833/2/258}{{\em \apj} {\bfseries
  833} (Dec., 2016) 258}, \href{http://arxiv.org/abs/1604.04625}{{\ttfamily
  arXiv:1604.04625 [astro-ph.HE]}}.

\bibitem{gralla-lupsasca-philippov2017}
S.~E. {Gralla}, A.~{Lupsasca}, and A.~{Philippov}, ``{Inclined Pulsar
  Magnetospheres in General Relativity: Polar Caps for the Dipole,
  Quadrudipole, and Beyond},''
  \href{http://dx.doi.org/10.3847/1538-4357/aa978d}{{\em \apj} {\bfseries 851}
  (Dec., 2017) 137}, \href{http://arxiv.org/abs/1704.05062}{{\ttfamily
  arXiv:1704.05062 [astro-ph.HE]}}.

\bibitem{komissarov2004}
S.~S. {Komissarov}, ``{Electrodynamics of black hole magnetospheres},''
  \href{http://dx.doi.org/10.1111/j.1365-2966.2004.07598.x}{{\em \mnras}
  {\bfseries 350} (May, 2004) 427--448}.

\bibitem{gralla-jacobson2014}
S.~E. {Gralla} and T.~{Jacobson}, ``{Spacetime approach to force-free
  magnetospheres},'' \href{http://dx.doi.org/10.1093/mnras/stu1690}{{\em
  \mnras} {\bfseries 445} (Dec., 2014) 2500--2534},
  \href{http://arxiv.org/abs/1401.6159}{{\ttfamily arXiv:1401.6159
  [astro-ph.HE]}}.

\bibitem{lyutikov-mckinney2011}
M.~{Lyutikov} and J.~C. {McKinney}, ``{Slowly balding black holes},''
  \href{http://dx.doi.org/10.1103/PhysRevD.84.084019}{{\em \prd} {\bfseries 84}
  no.~8, (Oct., 2011) 084019}, \href{http://arxiv.org/abs/1109.0584}{{\ttfamily
  arXiv:1109.0584 [astro-ph.HE]}}.

\bibitem{znajek1977}
R.~L. {Znajek}, ``{Black hole electrodynamics and the Carter tetrad},'' {\em
  \mnras} {\bfseries 179} (May, 1977) 457--472.

\bibitem{penna2015}
R.~F. {Penna}, ``{Energy extraction from boosted black holes: Penrose process,
  jets, and the membrane at infinity},''
  \href{http://dx.doi.org/10.1103/PhysRevD.91.084044}{{\em \prd} {\bfseries 91}
  no.~8, (Apr., 2015) 084044},
  \href{http://arxiv.org/abs/1503.00728}{{\ttfamily arXiv:1503.00728
  [astro-ph.HE]}}.

\bibitem{macdonald-thorne1982}
D.~{MacDonald} and K.~S. {Thorne}, ``{Black-hole electrodynamics - an
  absolute-space/universal-time formulation},'' {\em \mnras} {\bfseries 198}
  (Jan., 1982) 345--382.

\bibitem{penna2015impedance}
R.~F. {Penna}, ``{Black hole jet power from impedance matching},''
  \href{http://dx.doi.org/10.1103/PhysRevD.92.084017}{{\em \prd} {\bfseries 92}
  no.~8, (Oct., 2015) 084017},
  \href{http://arxiv.org/abs/1504.00360}{{\ttfamily arXiv:1504.00360
  [astro-ph.HE]}}.

\bibitem{TMN2008}
A.~{Tchekhovskoy}, J.~C. {McKinney}, and R.~{Narayan}, ``{Simulations of
  ultrarelativistic magnetodynamic jets from gamma-ray burst engines},''
  \href{http://dx.doi.org/10.1111/j.1365-2966.2008.13425.x}{{\em \mnras}
  {\bfseries 388} (Aug., 2008) 551--572},
  \href{http://arxiv.org/abs/0803.3807}{{\ttfamily arXiv:0803.3807}}.

\bibitem{michel1973mon}
F.~C. {Michel}, ``{Rotating Magnetospheres: an Exact 3-D Solution},''
  \href{http://dx.doi.org/10.1086/181169}{{\em \apjl} {\bfseries 180} (Mar.,
  1973) L133}.

\bibitem{contopoulos-kazanas-fendt1999}
I.~{Contopoulos}, D.~{Kazanas}, and C.~{Fendt}, ``{The Axisymmetric Pulsar
  Magnetosphere},'' \href{http://dx.doi.org/10.1086/306652}{{\em \apj}
  {\bfseries 511} (Jan., 1999) 351--358},
  \href{http://arxiv.org/abs/astro-ph/9903049}{{\ttfamily astro-ph/9903049}}.

\bibitem{gralla-wald2008}
S.~E. {Gralla} and R.~M. {Wald}, ``{A rigorous derivation of gravitational
  self-force},'' \href{http://dx.doi.org/10.1088/0264-9381/25/20/205009}{{\em
  Classical and Quantum Gravity} {\bfseries 25} no.~20, (Oct., 2008) 205009},
  \href{http://arxiv.org/abs/0806.3293}{{\ttfamily arXiv:0806.3293 [gr-qc]}}.

\bibitem{gralla-lupsasca-rodriguez2015}
S.~E. {Gralla}, A.~{Lupsasca}, and M.~J. {Rodriguez}, ``{Electromagnetic jets
  from stars and black holes},''
  \href{http://dx.doi.org/10.1103/PhysRevD.93.044038}{{\em \prd} {\bfseries 93}
  no.~4, (Feb., 2016) 044038},
  \href{http://arxiv.org/abs/1504.02113}{{\ttfamily arXiv:1504.02113 [gr-qc]}}.

\bibitem{numerical-recipes}
W.~H. {Press}, S.~A. {Teukolsky}, W.~T. {Vetterling}, and B.~P. {Flannery},
  {\em {Numerical recipes in FORTRAN. The art of scientific computing}}.
\newblock 1992.

\end{thebibliography}\endgroup

\end{document}